\documentclass[preprint,amsmath,amssymb,superscriptaddress,pre]{revtex4-1}
\usepackage{amssymb}
\usepackage{graphicx}
\usepackage{epstopdf}
\usepackage[bf,SL,BF]{subfigure}

\begin{document}
\title{Solitons as candidates for energy carriers in Fermi-Pasta-Ulam lattices}
\author{Yi Ming}
\email{meanyee@mail.ustc.edu.cn}
\author{Liu Ye}
\author{Han-Shuang Chen}
\affiliation{School of Physics and Material Science, Anhui University, Hefei, Anhui 230601, China}
\author{Shi-Feng Mao}
\affiliation{School of Nuclear Science and Technology, University of Science and Technology of China, Hefei, Anhui 230026, China}
\author{Hui-Min Li}
\affiliation{Supercomputing Center, University of Science and Technology of China, Hefei, Anhui 230026, China}
\author{Ze-Jun Ding}
\email{zjding@ustc.edu.cn}
\affiliation{Department of Physics and Hefei National Laboratory for Physical Sciences at the Microscale, University of Science and Technology of China, Hefei, Anhui 230026, China}
\affiliation{Key Laboratory of Strongly-coupled Quantum Matter Physics, Chinese Academy of Sciences}
\date{\today}

\begin{abstract}
Currently, effective phonons (renormalized or interacting phonons) rather than solitary waves (for short, solitons) are regarded as the energy carriers in nonlinear lattices. In this work, by using the approximate soliton solutions of the corresponding equations of motion and adopting the Boltzmann distribution for these solitons, the average velocities of solitons are obtained and are compared with the sound velocities of energy transfer. Excellent agreements with the numerical results and the predictions of other existing theories are shown in both the symmetric Fermi-Pasta-Ulam-$\beta$ lattices and the asymmetric Fermi-Pasta-Ulam-$\alpha \beta$ lattices. These clearly indicate that solitons are suitable candidates for energy carriers in Fermi-Pasta-Ulam lattices. In addition, the root-mean-square velocity of solitons can be obtained from the effective phonons theory.
\end{abstract}
\maketitle

\section{Introduction}
With the development of modern nanotechnology, energy transfer in low dimensional systems has attracted intense interest from fundamental research as well as applied research \cite{Pop2010, Li2012, Balandin2012266, Marconnet2013, Maldovan2013, Dhar2008, Lepri20031, Lepri2016}. One of the most intriguing phenomena is the anomalous energy transport in momentum-conserving lattices, which is the divergence of the thermal conductivity with the increasing lengths of lattices \cite{Dhar2008, Lepri20031, Lepri2016, Lepri19971896, Prosen20002857, Prosen2005, Narayan2002200601, Mai2006061202, Mai2007184301, Lee2005031202, Lee-Dadswell2008, Delfini2006060201, DelfiniP02007, Wang2004074302, Wang2004021204, Hu19982992, Hu20003828, Zhong2012060102, Zhang2013, Wang2013052112, Savin2014032102, Das2014204}. This has been experimentally verified in one dimensional (1D) carbon nanotubes \cite{Chang2008075903, Lee2017135901} and two dimensional (2D) graphenes \cite{Xu2014}.

Energy carriers in a linear (harmonic) lattice are phonons. The anomalous energy transport in it is attributed to the non-interaction of phonons. When nonlinear effects take place, the anomalous energy transport in 1D momentum-conserving lattices is mainly attributed to the L\'{e}vy walk of energy carriers on the microscopic level \cite{Denisov2003194301, Cipriani2005, Zaburdaev2011, Denisov2012031148, Zaburdaev2013, Zaburdaev2015, Liu2014, Dhar2013010103, Lepri2011030107}. On the mesoscopic level, the theory of nonlinear fluctuating hydrodynamics (NFH) is a powerful tool to study the anomalous energy transport in 1D momentum-conserving lattices \cite{Mendl2013, Mendl2014012147, Spohn2014, Das2014, van2012}. However, energy carriers in nonlinear lattices have not been identified explicitly so far. Because nonlinearity can cause interactions between phonons, the non-interaction of phonons are not anymore the reason for the anomalous energy transport in low dimensional nonlinear momentum-conserving lattices. Other than phonons, breathers can also be excited in nonlinear lattices \cite{Flach20081}. However, breather excitations in classical discrete lattices should be localized \cite{Flach20081, SZEFTEL2003215, Riseborough2012011129}. Breathers can thus only influence the energy transport by scattering the energy carriers \cite{Tsironis19996610, Gendelman20002381, Giardina20002144, Xiong2012020102, Xiong2014022117}.

Although the non-interacting phonons and breathers can be excluded from candidates for energy carriers in nonlinear lattices, there is still a debate about whether the energy carriers are solitary waves \cite{Fillipov1998, Theodorakopoulos19992293, Zhang20003541, Hu20003828, Aoki20014029, Li2005, Zhao2005025507, Zhao2006140602} (for simplicity, we refer them as solitons in this work) or effective phonons \cite{Li2010054102,Li2012AIP, Li2013, Liu2014174304, Liu2015}. Solitons are localized waves that propagate over long distances almost without deformation even after collisions with phonons or solitons \cite{Dauxois2006Physics}. They can be expressed as the combinations of linear waves \cite{McMillan19761496, Abe1979, ABE1980202, Truskinovsky2014042903}. More than fifty years ago, by reducing the Fermi-Pasta-Ulam (FPU) problem \cite{FermiStudies} to the Korteweg-de Vries (KdV) equation based on the long-wavelength approximation, Zabusky and Kruskal coined the name ``soliton'' and numerically discovered that the initial cosine wave can break up into nine solitons with different speeds \cite{Zabusky1965240}. This phenomenon of multisoliton fission has been experimentally proved in shallow water recently \cite{Trillo2016144102}. Remarkably, the solitons have also been extracted from the measured data of the shallow water ocean surface waves \cite{Costa2014108501}. Besides in the continuous nonlinear systems, the existence of solitons in FPU lattices has been proved \cite{Friesecke1994, Friesecke1999, FRIESECKE2002211, SMETS1997266, Iooss2000}. A series of numerical works have directly observed solitons in FPU lattices \cite{PEYRARD1986268, Flytzanis1989, Kosevich19932058, *Kosevich2004, Archilla2015022912, Szeftel19993982, SZEFTEL2000225, Neogi2008064306, Jin2010037601, Truskinovsky2014042903}. Therefore, solitons were naturally supposed to be the energy carriers in the nonlinear momentum-conserving lattices \cite{Fillipov1998, Theodorakopoulos19992293, Zhang20003541, Hu20003828, Aoki20014029, Li2005, Zhao2005025507, Zhao2006140602}. It is also noteworthy that the first experiment of the nanoscale thermal rectification was also speculated as the result of asymmetric transport of solitons \cite{Chang06}.

However, by comparing the numerical results of the sound velocities of energy transfer with the predictions by solitons theory and effective phonons theory (EPT), it is obtained that energy carriers in the FPU-$\beta$ lattices are effective phonons (which are regarded as the interacting phonons) but solitons are ``ruled out'' \cite{Li2010054102}. Since then, effective phonons \cite{Li2012AIP, Li2013, Liu2014174304, Liu2015} are regarded as the energy carriers in nonlinear lattices.

It is noticed that the numerical results of sound velocities were compared only with the velocity of one specific soliton in Ref. [\onlinecite{Li2010054102}]. However, at a thermal equilibrium state or a non-equilibrium steady state, there should exist a statistical distribution of solitons with different velocities rather than one soliton with a specific velocity. Therefore, the role of the solitons in nonlinear momentum-conserving lattices should be further investigated. In addition, EPT cannot predict accurately the sound velocities in nonlinear lattices with asymmetric interparticle interactions (including FPU-$\alpha\beta$ lattices) \cite{Zhang2013, Liu2014174304, Liu2015}. Where strong finite-size effects make the thermal conductivity saturate to a finite value for a wide range of lattice lengths. Nevertheless, it will be divergent in the thermodynamic limit \cite{Zhong2012060102, Zhang2013, Wang2013052112, Savin2014032102, Das2014204}. Even the optimal effective phonon theory based on variational approach (vEPT) cannot predict the accurate sound velocities in lattices with very strong asymmetric interparticle interactions either \cite{Liu2015}. Therefore, energy carriers in FPU lattices should be revisited.

Same as in Refs. [\onlinecite{Aoki20014029, Li2010054102}], we compare our results with the sound velocities of energy transfer to study the energy carriers in this work. We approximately obtain the soliton solutions of the equations of motion of FPU lattices by using weak formulation \cite{DUNCAN1992505}. Adopting further the Boltzmann distribution for these solitons, the average velocities of solitons are obtained. The relation with the EPT is also provided. According to the EPT, by projecting the dynamics of the FPU lattices onto a soliton, the root-mean-square velocity of solitons can be obtained. For comparison, we also numerically compute the sound velocities in FPU lattices. The equations of motion are integrated by using implicit midpoint algorithm \cite{Ming2016} with periodic boundary conditions. Sound velocity can be obtained by calculating the ratio of the lowest peak frequency of the power spectrum to the corresponding harmonic phonon frequency \cite{Zhang2013, Liu2015}. In the FPU-$\beta$ lattices, the average velocities of solitons agree well with the predictions of EPT. The numerical results agree well also with the predictions of EPT. This confirms the validity of the numerical program. For FPU-$\alpha\beta$ lattices, EPT, even vEPT cannot predict the sound velocities accurately. However, the average velocities of solitons agree well with the predictions of NFH as well as the numerical results. Therefore, we conclude that solitons are suitable candidates for energy carriers in FPU lattices.

The rest of the paper is organized as follows. In Sec. \ref{sec2-1}, the model of the studied nonlinear lattices is presented. The approximate soliton solutions of the equations of motion are obtained in Sec. \ref{sec2-2}. How to calculate the average velocity of solitons and its relation with the EPT are presented in Sec. \ref{sec2-3}. The results are presented in Sec. \ref{sec3}. Finally, the conclusion and the discussion are presented in Sec. \ref{sec4}.

\section{Model and methods}\label{sec2}
\subsection{Model}\label{sec2-1}
The dimensionless Hamiltonian of 1D momentum-conserving lattice is
\begin{equation}\label{ham}
H=\sum_j[\frac{\dot{u}_j^2}{2}+V(u_j-u_{j-1})],
\end{equation}
where $u_j$ is the displacement of the $j$th particle from its equilibrium position, the dot denotes the time derivative. The corresponding equation of motion can be expressed as
\begin{equation}\label{eom1}
\ddot{u}_j=V^\prime (u_{j+1}-u_j)-V^\prime (u_j-u_{j-1}),
\end{equation}
where the prime denotes the derivative of the function with respect to its argument.
It can be reexpressed as
\begin{equation}\label{eom}
\ddot{\phi}_j=V^\prime (\phi_{j+1})-2V^\prime (\phi_{j})+V^\prime (\phi_{j-1}),
\end{equation}
where $\phi_j=u_j-u_{j-1}$ denotes the relative displacement between the adjacent particles.

The potential of the FPU lattice can be expressed as
\begin{equation}\label{pot}
V(\phi_j)=\frac{1}{2}\phi_j^2+\frac{\alpha}{3}\phi_j^3+\frac{\beta}{4}\phi_j^4.
\end{equation}
The lattice is respectively called FPU-$\beta$ lattice or FPU-$\alpha$ lattice for $\alpha = 0$ or $\beta=0$. Otherwise, it is called FPU-$\alpha\beta$ lattice. In this work, we choose $\beta=1$ and $0\le \alpha \le 2$ as in Refs. [\onlinecite{Zhang2013, Liu2015}]. Same results can be obtained for $\alpha<0$. To compare with the results of Ref. [\onlinecite{Li2010054102}], the nonlinear lattices with
\begin{equation}
V(\phi_j)=\frac{|\phi|^k}{k}, \qquad k=3,4,\text{or} \,5
\end{equation}
are also studied in the subsection \ref{sec3-2}.

\subsection{The approximate soliton solutions}\label{sec2-2}
The equation of motion Eq. \eqref{eom} permits the bell-shaped soliton solutions for $\phi_j$ (corresponding to the kink-shaped soliton solutions for $u_j$) \cite{PEYRARD1986268, Flytzanis1989, Kosevich19932058, *Kosevich2004, Archilla2015022912, Szeftel19993982, SZEFTEL2000225, Neogi2008064306, Jin2010037601, Truskinovsky2014042903}. Soliton is a travelling wave solution and can be expressed formally as
\begin{equation}\label{tra}
\phi_j(t)=\phi(j-ct)\equiv \phi(z),
\end{equation}
where $c$ is its velocity. Substituting it into Eq. \eqref{eom}, a differential-difference equation can be obtained as
\begin{equation}\label{dde}
c^2\phi^{\prime\prime}(z)=V^\prime [\phi(z+1)]-2V^\prime [\phi(z)]+V^\prime [\phi(z-1)].
\end{equation}
It can be approximately solved to obtain the soliton solutions by using the weak formulation \cite{DUNCAN1992505}.

One should recall that the equation of motion of the FPU-$\beta$ lattices and the FPU-$\alpha$ lattices will result in the modified KdV equation and the KdV equation in the long-wavelength approximations \cite{ZABUSKY1981195}. The corresponding soliton solutions are sech-shaped and sech-squared-shaped respectively. We thus express the ansatz soliton solution as a sech-shaped function
\begin{equation}\label{soliton}
\phi(z)=A\, \text{sech}(q z),
\end{equation}
where $A$ and $q$ can be determined from the following weak formulation Eq. \eqref{wfe}. The half-height width of the soliton is $W={2\cosh^{-1}(2)}/{q}$, where $q$ can be treated as its wave vector \cite{Neogi2008064306} and $\lambda=2\pi/q$ is the corresponding wavelength. The amplitude $A<0$ corresponds to the compressional soliton and $A>0$ to the rarefaction (dilatational) soliton. Wave vectors $\pm q$ correspond to the same soliton. We have also checked the sech-squared-shaped solitons with $\phi(z)=A\, \text{sech}^2(q z)$, the results (not shown here) are the same except for the effective widths of solitons (defined below).

Multiplying Eq. \eqref{dde} by a trial function $\psi(z)$ and integrating it over the whole real axis, the weak formulation is obtained. It can be transformed into
\begin{eqnarray}\label{wfe}
&&c^2\int_{-\infty}^\infty \psi^{\prime\prime}(z)\phi(z)dz\nonumber\\
&&=\int_{-\infty}^\infty[\psi(z+1)-2\psi(z)+\psi(z-1)] V^\prime[\phi(z)]dz
\end{eqnarray}
by integrating by parts on the left-hand side and changing the variable of integration on the right-hand side as well as assuming that $\psi$, $\phi$ and their derivatives behave reasonably as $|z|\rightarrow \infty$.

As in Ref. [\onlinecite{DUNCAN1992505}], we choose $\psi(z)=z^2$ and $z^4$ in Eq. \eqref{wfe}. Substituting the ansatz solution Eq. \eqref{soliton} into Eq. \eqref{wfe}, it is obtained that
\begin{equation}\label{amp}
A_{\pm}=\frac{-2 \alpha \pm\sqrt{4 \alpha^2 +2(c^2-1)\pi^2}}{\pi}
\end{equation}
and
\begin{eqnarray}\label{vec}
q_{\pm}=\sqrt{\frac{2}{\pi c^2}\Big[6\pi (c^2-1)
+\alpha (\pi^2-12)A_{\pm}\Big]},
\end{eqnarray}
where the subscripts ``$+$'' and ``$-$'' correspond to the plus and minus signs in Eq. \eqref{amp} respectively. When $\alpha=0$, the amplitude obtained from Eq. \eqref{amp} coincides with the results in Refs. [\onlinecite{Zhang20003541, Li2010054102}]. Once the soliton solution is obtained, its total kinetic energy can be calculated according to Ref. [\onlinecite{Zhang20003541}] as
\begin{equation}\label{tke}
E_k={c^2}\int_{-\infty}^\infty \frac{\phi^2(z)}{2} dz=\frac{A^2 c^2}{q}.
\end{equation}

\subsection{Average velocity of solitons}\label{sec2-3}
To calculate the average velocity of solitons, the statistical distribution of solitons has to be determined in advance. Although the statistical mechanics of some nonlinear integrable (soliton-bearing) systems have already been developed \cite{Currie1980477, Theodorakopoulos1984871, Fratalocchi2008044101}, the statistical distribution of solitons in nonlinear nonintegrable lattices has not yet been reported so far to our knowledge. In this work, the Boltzmann distribution is used for solitons of the FPU lattices in which the existence of solitons has been proved \cite{Friesecke1994, Friesecke1999, FRIESECKE2002211, SMETS1997266, Iooss2000}. Same as the other kind of nonlinear excitations which are referred to as discrete breathers (intrinsic localized modes) \cite{Sievers1988970, Ming2017}, solitons in FPU lattices can be also regarded as a dilute gas of nonlinear defects. There thus exist an activation energy $\epsilon$ with respect to the number of solitons. The number of solitons is accordingly assumed to be
\begin{equation}\label{disfun}
f(\epsilon) =N_0 \exp[-\frac{\epsilon}{k_B T}],
\end{equation}
where $N_0$ is a constant relevant to the length of lattice, $k_B$ is the Boltzmann constant and $T$ is the temperature.

To determine the activation energy $\epsilon$, we briefly recall the effective phonons theory first. By using the Zwanzig-Mori projection formalism \cite{Lepri19987165, Lee20133237, Ming2017}, the dynamics of the nonlinear lattice Eq. \eqref{ham} in thermal equilibrium can be projected onto one of its harmonic normal modes. The corresponding renormalized frequency can be given by
\begin{equation}\label{eq13}
\Omega_l^2=\frac{\langle \dot{Q}_l^2\rangle}{\langle {Q}_l^2\rangle}=\frac{\langle \frac{1}{2}\dot{Q}_l^2\rangle}{\langle \frac{1}{2}\omega_l^2 {Q}_l^2\rangle}\omega_l^2=\frac{\langle K_l\rangle}{\langle U_l\rangle}\omega_l^2,
\end{equation}
where $\omega_l$, $Q_l$, $K_l$ and $U_l$ are respectively the normal-mode frequency, the amplitude, the kinetic energy and the harmonic potential energy of the $l$-th normal mode. $\langle \cdot\rangle$ denotes the ensemble average with respect to the single site probability density \cite{Spohn2014, Liu2015}
\begin{equation}\label{eq14}
\rho_s(v_j,\phi_j)=\frac{1}{Z}\exp\Big\{-\frac{1}{k_BT}\Big[\frac{v_j^2}{2}+V(\phi_j)+P\phi_j\Big]\Big\},
\end{equation}
where $Z$ is the corresponding partition function, $P$ is the pressure, $v_j=\dot{u}_j$ and $\phi_j=u_j-u_{j-1}$ are the single site variables.
 Based on the effective phonons theory \cite{Li2010054102, Liu2015}, the sound velocity can be expressed as
\begin{equation}\label{eq15}
  c_s=\sqrt{\frac{\langle K_l\rangle}{\langle U_l\rangle}}.
\end{equation}

We regard the solitons as the nonlinear eigen-modes of the nonlinear lattices and thus project the dynamics onto one soliton. By substituting the kinetic energy (Eq. \eqref{tke}) and the harmonic potential energy of the soliton $\int_{-\infty}^\infty {\phi^2(z)} dz/{2}$ into Eq. \eqref{eq15}, the sound velocity can be expressed as
\begin{equation}\label{eq16}
  c_s=\sqrt{\frac{\langle E_k\rangle}{\langle \int_{-\infty}^\infty {\phi^2(z)} dz/{2}\rangle}}= \sqrt{\frac{\langle E_k\rangle}{\langle E_k/c^2\rangle}} \approx \sqrt{\langle c^2\rangle}.
\end{equation}
This is just the root-mean-square velocity of solitons.

In Eq. \eqref{eq16}, $c_s$ can be regarded as depending only on the kinetic energy of solitons, we thus suppose that the ensemble average can be taken with respect to
\begin{equation}\label{veldis}
\rho_v=\frac{1}{\sqrt{2\pi k_BT}}\exp\Big[-\frac{{v_j^2}}{2k_BT}\Big],
\end{equation}
where $v_j^2/2$ is the kinetic energy of a single site. Comparing Eq. \eqref{disfun} with Eq. \eqref{veldis}, the activation energy $\epsilon$ can be chosen as the kinetic energy of a single site. For a soliton, $\epsilon$ should be expressed as the average kinetic energy per site of it
\begin{equation}\label{eq19}
\epsilon=\frac{E_k}{W_e}=\frac{E_k}{2\eta W},
\end{equation}
where $W_e$ is the effective width of the soliton and $\eta$ is a fitting parameter to determine $W_e$. As shown in the following, $W_e$ is independent on $\eta$ and can be chosen as the wavelength $\lambda$ of the soliton for FPU-$\beta$ lattice. This can be attributed to that the ansatz solution Eq. \eqref{soliton} approaches the exact soliton solution for FPU-$\beta$ lattice. Because the FPU-$\beta$ lattice will result in the modified KdV equation in the long-wavelength approximations \cite{ZABUSKY1981195}. The soliton solution of the modified KdV equation is sech-shaped. This coincides with our ansatz solution Eq. \eqref{soliton}. We expect that the effective width of a soliton $W_e$ is just its wavelength $\lambda$ when the soliton solution is exact.

Considering the potential degeneracy between the rarefaction solitons and the compressional solitons, the average velocity of solitons can thus be expressed by neglecting the interactions between them as
\begin{equation}\label{avvel}
c_s=\frac{\int_0^\infty [(1-\xi)c_r +\xi c_c] \exp(-\frac{\epsilon}{k_B T})d\epsilon }{\int_0^\infty \exp(-\frac{\epsilon}{k_B T})d\epsilon},
\end{equation}
where $c_r$ and $c_c$ are respectively the velocities of rarefaction and compressional solitons correspond to the same $\epsilon$, $\xi$ is the excitation probability of a compressional soliton.

\section{Results}\label{sec3}

\subsection{The FPU-$\beta$ lattice}\label{sec3-1}
\begin{center}
\begin{figure}[htbp]
\includegraphics[width=0.5\textwidth]{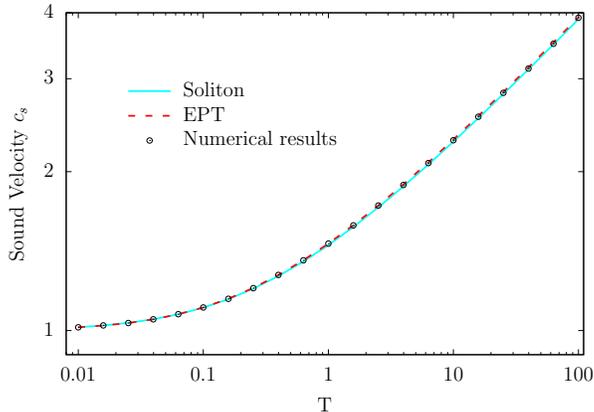}
\caption{\label{fig1} Sound velocity $c_s$ as the function of temperature $T$ for the 1D FPU-$\beta$ lattice. The solid line corresponds to the average velocities of solitons and the dashed line corresponds to the predictions of EPT. The symbols correspond to the numerical results.}
\end{figure}
\end{center}

From Eq. \eqref{amp}, the amplitudes of the rarefaction solitons and the compressional solitons in the FPU-$\beta$ lattices are obtained as $A_+=-A_-=\sqrt{2(c^2-1)}$. This can be attributed to the symmetry of the FPU-$\beta$ potential. According to Eq. \eqref{tke}, the velocities of these degenerate solitons (two kinds of solitons correspond to the same $\epsilon$) are equal to each other. We thus expect that these two kinds of solitons can be excited with the equal probability $\xi=1/2$. However, one should notice that $c_s$ is actually independent on $\xi$ when $c_r=c_c$ (see Eq. \eqref{avvel}) and thus the following results of FPU-$\beta$ lattices are independent on $\xi$.

The average velocities of solitons can be obtained according to Eq. \eqref{avvel}. The results are depicted in Fig. \ref{fig1} as a function of the temperature $T$. The effective width is $W_e=\lambda$. It should be emphasized here that the results are fitting-parameter-free when considering the aforementioned independence of $c_s$ on $\xi$. Comparing with the predictions of EPT, an excellent agreement is obtained. Because EPT has accurately predicted the sound velocities of FPU-$\beta$ lattices \cite{Li2010054102}, this agreement clearly indicates that solitons are indeed candidates for energy carriers in FPU-$\beta$ lattices. In Fig. \ref{fig1}, the numerical results are also depicted as symbols and thus confirm the validity of our numerical program.

We should mention here that our results are negligibly lower than the predictions of EPT at high temperature (the relative discrepancy is less than $0.9\%$ at $T=100$). This can be attributed to that the sech-shaped soliton is not completely accurate at high temperature where the long-wavelength approximations are not suitable anymore \cite{FRIESECKE2002211, Truskinovsky2014042903}. However, if we choose $\eta=1.2358$ corresponding to $W_e\approx 6.51/q$ rather than $W_e=\lambda=2\pi/q$, the discrepancies are almost removed at high temperatures. At low temperature region, $\epsilon$ and $c_s$ are not very sensitively dependent on the width of the soliton because the soliton is very wide and its total kinetic energy is very low.

\subsection{The nonlinear lattice with $V(\phi_j)=|\phi_j|^{k}/k$}\label{sec3-2}
\begin{center}
\begin{figure}[htbp]
\includegraphics[width=0.5\textwidth]{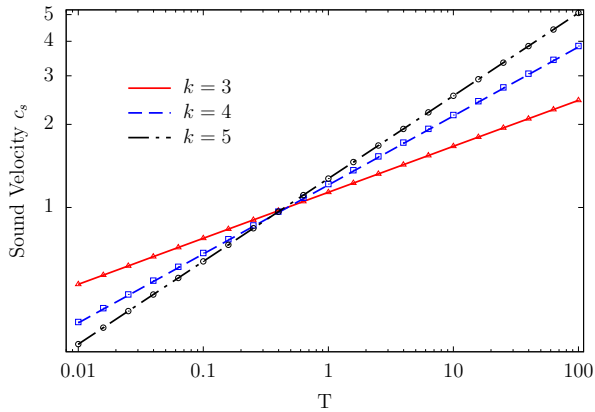}
\caption{\label{fig2} Sound velocity $c_s$ as the function of temperature $T$ for the 1D nonlinear lattices with $V(\phi_j)=|\phi_j|^{k}/k$, where $k=3$, $4$ and $5$. The symbols correspond to the predictions of EPT and the lines are the average velocities of solitons.}
\end{figure}
\end{center}

For comparison, we also study the sound velocity in the nonlinear lattices whose nearest-neighbor interaction potentials are $V(\phi_j)=|\phi_j|^{k}/k$ with $k \ge 3$. Solitons can be excited in these nonlinear lattices \cite{Friesecke1994}. It should be emphasized here that the amplitudes of solitons, $A$'s, are always positive when the potentials are $V(\phi_j)=\phi_j^{k}/k$ with the odd $k$'s. This indicates that the rarefaction solitons are permitted only in the nonlinear lattices with the asymmetric cubic and quintic interaction potentials. When the potential is symmetric (e.g. $V(\phi_j)=|\phi_j|^{k}/k$), compressional solitons are also permitted with the same probability. Therefore, the average velocities of solitons can be calculated by using Eq. \eqref{avvel} with $\xi=1/2$ also. The results are shown in Fig. \ref{fig2} with the fitting parameter $\eta$ equals to $1.5695$ and $1.0765$ for $k=3$ and $5$ respectively. Same as the FPU-$\beta$ lattices, the results for $k=4$ are also fitting-parameter-free with $W_e=\lambda$. In Fig. \ref{fig2}, the predictions of EPT are plotted as symbols for clarity. The excellent agreements between them and our results are obtained. This indicates that solitons are candidates for energy carriers in the nonlinear lattices with $V(\phi_j)=|\phi_j|^{k}/k$.

\subsection{The FPU-$\alpha \beta$ lattice}\label{sec3-3}
\begin{center}
\begin{figure}[htbp]
\includegraphics[width=0.5\textwidth]{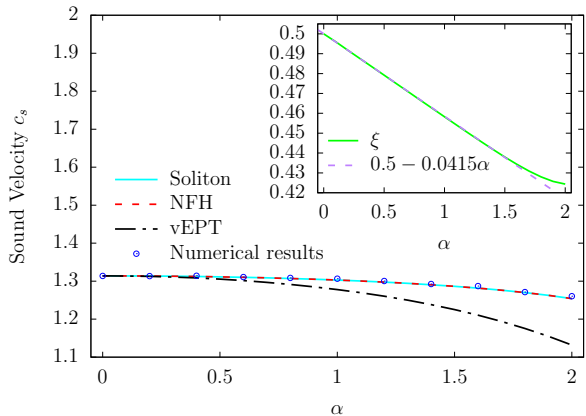}
\caption{\label{fig3}  Sound velocity $c_s$ as the function of $\alpha$ for the 1D FPU-$\alpha\beta$ lattices at $T=0.5$. The solid line corresponds to the average velocities of solitons. The dashed line corresponds to the predictions of NFH and the dashed-dotted line corresponds to the predictions of vEPT. The symbols correspond to the numerical results. The fitted excitation probabilities $\xi$ are plotted in the inset as the solid line. It coincides with the function $0.5-0.0415\alpha$ when $\alpha <1.5$.}
\end{figure}
\end{center}

To further confirm that solitons are candidates for energy carriers in FPU lattices, average velocities of solitons in FPU-$\alpha \beta$ lattices are calculated. Because of the asymmetry of the potential, velocities of two degenerate solitons are not the same anymore. All rarefaction solitons are supersonic with $c>1$ same as in the FPU-$\beta$ lattices. But compressional solitons with their amplitudes $-4\alpha/\pi<A<0$ are subsonic solitons with $c<1$.

We should mention here that there exist two specific $\epsilon$, $\epsilon_1$ and $\epsilon_2$, with $\epsilon_1<\epsilon_2$. Here $\epsilon_1$ corresponds to $A_+=A_-=-2\alpha/\pi$ and $\epsilon_2$ to $q_-=0$. In the region $0<\epsilon<\epsilon_1$, $A_+$ is a double-valued function of $\epsilon$. One corresponds to a rarefaction soliton with $A_+>0$, the other to a compressional soliton with $A_+<0$. In the region $\epsilon_1<\epsilon<\epsilon_2$, $A_-$ is the amplitude of a compressional soliton. However, in these two energy regions, the wave vector $q$ of the compressional soliton is purely imaginary. Thus the ansatz solution Eq. \eqref{soliton} transforms to a singular periodic solution $A\sec(|q| z)$. The singular periodic solution does indeed exist for KdV equation \cite{Chen1979264}. It leads to a positive eigenvalue of the associated linear spectrum problem while solitons lead to negative eigenvalues in the inverse scattering method for solving the KdV equation. Therefore, the singular periodic solution is often coined as (zeroth-order) ``positon'' \cite{Rasinariu1996, Matveev2002}. However, potentially because of lack of direct physical interpretation, the singular solution is mainly interested in mathematically oriented studies.

In this work, we attribute this singular periodic solution to that the ansatz solution Eq. \eqref{soliton} is not accurate for FPU-$\alpha\beta$ lattices. Because $q$ becomes real in aforementioned two energy regions when we choose $\psi(z)=z^6$ in the weak formulation Eq. \eqref{wfe} which should be satisfied by the exact soliton solution for any reasonable choice of $\psi(z)$. There thus exist compressional solitons in these two energy regions. The obtained soliton solutions are used to achieve the following results. However, because $\epsilon$ in Eq. \eqref{eq19} is independent on $q$, there is no difference in the results obtained by directly using the singular periodic solutions. Therefore, whether the singular periodic solutions can be exist in FPU-$\alpha \beta$ lattices deserve further studies.

\begin{center}
\begin{figure}[htbp]
\includegraphics[width=0.5\textwidth]{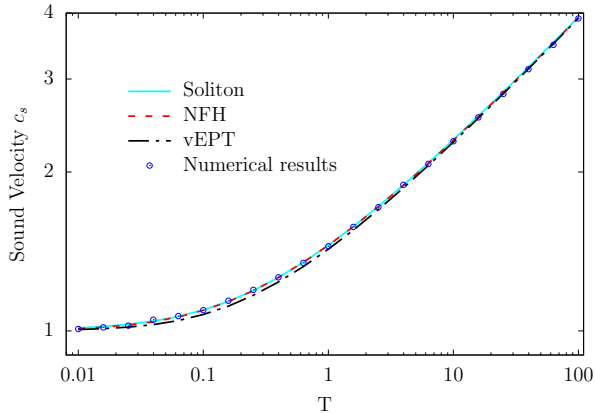}
\caption{\label{fig4} Sound velocity $c_s$ as the function of temperature $T$ for the 1D FPU-$\alpha\beta$ lattices with $\alpha=1$. The solid line corresponds to the average velocities of solitons. The dashed line corresponds to the predictions of NFH and the dashed-dotted line corresponds to the predictions of vEPT. The symbols correspond to the numerical results.}
\end{figure}
\end{center}

Excitation probability of the compressional soliton $\xi$ should be determined first to calculate the average velocity. When $\alpha > 0$, according to the aforementioned results of the nonlinear lattices with $V(\phi_j)=\phi_j^{3}/3$, we expect that the rarefaction solitons should be excited with higher probability than the compressional solitons. It is thus expected that $\xi<1/2$ and decreases with $\alpha$. However, we can only obtain $\xi$ by fitting our results to the predictions of NFH. Results are shown in Fig. \ref{fig3} with $\eta=1.2358$ which has been used for $\alpha=0$. As it should be, our results recover the predictions of NFH. All of them agree very well with the numerical results. As shown in Ref. [\onlinecite{Liu2015}], vEPT cannot predict accurately the sound velocities. The curve of $\xi$ is shown in the inset of Fig. \ref{fig3}. It does indeed decrease with $\alpha$. When $\alpha<1.5$, the dependence of $\xi$ on $\alpha$ is linear and can be fitted by a function $0.5-0.0415\alpha$. We speculate that $\eta=1.2358$ is not suitable when $\alpha>1.5$ and this deserve further study. However, the excellent agreements in  Fig. \ref{fig3} indicate that solitons are still candidates for energy carriers in FPU-$\alpha\beta$ lattices.

To further confirm this conclusion, the sound velocity as a function of temperature for the 1D FPU-$\alpha\beta$ lattices with $\alpha=1$ is studied by using the corresponding $\xi$ obtained from the inset of Fig. \ref{fig3}. The results are shown in Fig. \ref{fig4}. Our results recover the predictions of NFH again. All of them agree very well with the numerical results. It is thus obtained that $\xi$ is independent on the temperature $T$. However, there are slight discrepancies between the predictions of vEPT and the numerical results.

\section{Conclusion and discussion}\label{sec4}
In summary, by using the approximate soliton solutions of the equations of motion and using the Boltzmann distribution for solitons, the average velocities of solitons in FPU lattices are obtained. The results agree excellently with the sound velocities of energy transfer not only in the symmetric FPU-$\beta$ lattices but also in the asymmetric FPU-$\alpha\beta$ lattices. We thus conclude that solitons are still candidates for energy carriers in FPU lattices.

By using the Zwanzig-Mori projection formalism to project the dynamics of the nonlinear lattices onto one soliton solution, the root-mean-square velocity of the soliton is obtained based on the effective phonons theory. The relations between solitons and the effective phonons deserve further study.

The anomalous energy transport in FPU lattices can be attributed to the properties of solitons which can almost conserve their identities after collisions. Our results also confirm the speculation of Ref. [\onlinecite{Chang06}] that thermal rectification can be the result of asymmetric transport of solitons. However, because there is a statistical distribution of solitons with different velocities, solitons cannot be detected by visualizing the spatiotemporal evolutions of local energy densities or relative displacements. We hope that our investigate can motivate further study on energy transfer in nonlinear lattices from the point of view of solitons, especially on the relations between solitons and thermal rectification.

\begin{acknowledgments}
We thank the referees for their constructive comments. We also thank Profs. Hong Zhao, Yong Zhang (Xiamen University) and Tao Jin (Shaanxi Normal University) for their insightful comments on the first version of this manuscript. Z.-J.D. is supported by the National Natural Science Foundation of China (Grant No. 11574289). Some numerical calculations in this work were performed on the supercomputing system in the Supercomputing Center of USTC.
\end{acknowledgments}
%

\begin{thebibliography}{101}%
\makeatletter
\providecommand \@ifxundefined [1]{%
 \@ifx{#1\undefined}
}%
\providecommand \@ifnum [1]{%
 \ifnum #1\expandafter \@firstoftwo
 \else \expandafter \@secondoftwo
 \fi
}%
\providecommand \@ifx [1]{%
 \ifx #1\expandafter \@firstoftwo
 \else \expandafter \@secondoftwo
 \fi
}%
\providecommand \natexlab [1]{#1}%
\providecommand \enquote  [1]{``#1''}%
\providecommand \bibnamefont  [1]{#1}%
\providecommand \bibfnamefont [1]{#1}%
\providecommand \citenamefont [1]{#1}%
\providecommand \href@noop [0]{\@secondoftwo}%
\providecommand \href [0]{\begingroup \@sanitize@url \@href}%
\providecommand \@href[1]{\@@startlink{#1}\@@href}%
\providecommand \@@href[1]{\endgroup#1\@@endlink}%
\providecommand \@sanitize@url [0]{\catcode `\\12\catcode `\$12\catcode
  `\&12\catcode `\#12\catcode `\^12\catcode `\_12\catcode `\%12\relax}%
\providecommand \@@startlink[1]{}%
\providecommand \@@endlink[0]{}%
\providecommand \url  [0]{\begingroup\@sanitize@url \@url }%
\providecommand \@url [1]{\endgroup\@href {#1}{\urlprefix }}%
\providecommand \urlprefix  [0]{URL }%
\providecommand \Eprint [0]{\href }%
\providecommand \doibase [0]{http://dx.doi.org/}%
\providecommand \selectlanguage [0]{\@gobble}%
\providecommand \bibinfo  [0]{\@secondoftwo}%
\providecommand \bibfield  [0]{\@secondoftwo}%
\providecommand \translation [1]{[#1]}%
\providecommand \BibitemOpen [0]{}%
\providecommand \bibitemStop [0]{}%
\providecommand \bibitemNoStop [0]{.\EOS\space}%
\providecommand \EOS [0]{\spacefactor3000\relax}%
\providecommand \BibitemShut  [1]{\csname bibitem#1\endcsname}%
\let\auto@bib@innerbib\@empty
\bibitem [{\citenamefont {Pop}(2010)}]{Pop2010}%
  \BibitemOpen
  \bibfield  {author} {\bibinfo {author} {\bibfnamefont {E.}~\bibnamefont
  {Pop}},\ }\href {\doibase 10.1007/s12274-010-1019-z} {\bibfield  {journal}
  {\bibinfo  {journal} {Nano Res.}\ }\textbf {\bibinfo {volume} {3}},\ \bibinfo
  {pages} {147} (\bibinfo {year} {2010})}\BibitemShut {NoStop}%
\bibitem [{\citenamefont {Li}\ \emph {et~al.}(2012)\citenamefont {Li},
  \citenamefont {Ren}, \citenamefont {Wang}, \citenamefont {Zhang},
  \citenamefont {H\"anggi},\ and\ \citenamefont {Li}}]{Li2012}%
  \BibitemOpen
  \bibfield  {author} {\bibinfo {author} {\bibfnamefont {N.}~\bibnamefont
  {Li}}, \bibinfo {author} {\bibfnamefont {J.}~\bibnamefont {Ren}}, \bibinfo
  {author} {\bibfnamefont {L.}~\bibnamefont {Wang}}, \bibinfo {author}
  {\bibfnamefont {G.}~\bibnamefont {Zhang}}, \bibinfo {author} {\bibfnamefont
  {P.}~\bibnamefont {H\"anggi}}, \ and\ \bibinfo {author} {\bibfnamefont
  {B.}~\bibnamefont {Li}},\ }\href {\doibase 10.1103/RevModPhys.84.1045}
  {\bibfield  {journal} {\bibinfo  {journal} {Rev. Mod. Phys.}\ }\textbf
  {\bibinfo {volume} {84}},\ \bibinfo {pages} {1045} (\bibinfo {year}
  {2012})}\BibitemShut {NoStop}%
\bibitem [{\citenamefont {Balandin}\ and\ \citenamefont
  {Nika}(2012)}]{Balandin2012266}%
  \BibitemOpen
  \bibfield  {author} {\bibinfo {author} {\bibfnamefont {A.~A.}\ \bibnamefont
  {Balandin}}\ and\ \bibinfo {author} {\bibfnamefont {D.~L.}\ \bibnamefont
  {Nika}},\ }\href {\doibase http://dx.doi.org/10.1016/S1369-7021(12)70117-7}
  {\bibfield  {journal} {\bibinfo  {journal} {Mater. Today}\ }\textbf {\bibinfo
  {volume} {15}},\ \bibinfo {pages} {266 } (\bibinfo {year}
  {2012})}\BibitemShut {NoStop}%
\bibitem [{\citenamefont {Marconnet}\ \emph {et~al.}(2013)\citenamefont
  {Marconnet}, \citenamefont {Panzer},\ and\ \citenamefont
  {Goodson}}]{Marconnet2013}%
  \BibitemOpen
  \bibfield  {author} {\bibinfo {author} {\bibfnamefont {A.~M.}\ \bibnamefont
  {Marconnet}}, \bibinfo {author} {\bibfnamefont {M.~A.}\ \bibnamefont
  {Panzer}}, \ and\ \bibinfo {author} {\bibfnamefont {K.~E.}\ \bibnamefont
  {Goodson}},\ }\href {\doibase 10.1103/RevModPhys.85.1295} {\bibfield
  {journal} {\bibinfo  {journal} {Rev. Mod. Phys.}\ }\textbf {\bibinfo {volume}
  {85}},\ \bibinfo {pages} {1295} (\bibinfo {year} {2013})}\BibitemShut
  {NoStop}%
\bibitem [{\citenamefont {{Maldovan}}(2013)}]{Maldovan2013}%
  \BibitemOpen
  \bibfield  {author} {\bibinfo {author} {\bibfnamefont {M.}~\bibnamefont
  {{Maldovan}}},\ }\href {\doibase 10.1038/nature12608} {\bibfield  {journal}
  {\bibinfo  {journal} {\nat}\ }\textbf {\bibinfo {volume} {503}},\ \bibinfo
  {pages} {209} (\bibinfo {year} {2013})}\BibitemShut {NoStop}%
\bibitem [{\citenamefont {Dhar}(2008)}]{Dhar2008}%
  \BibitemOpen
  \bibfield  {author} {\bibinfo {author} {\bibfnamefont {A.}~\bibnamefont
  {Dhar}},\ }\href {\doibase 10.1080/00018730802538522} {\bibfield  {journal}
  {\bibinfo  {journal} {Adv. Phys.}\ }\textbf {\bibinfo {volume} {57}},\
  \bibinfo {pages} {457} (\bibinfo {year} {2008})}\BibitemShut {NoStop}%
\bibitem [{\citenamefont {Lepri}\ \emph {et~al.}(2003)\citenamefont {Lepri},
  \citenamefont {Livi},\ and\ \citenamefont {Politi}}]{Lepri20031}%
  \BibitemOpen
  \bibfield  {author} {\bibinfo {author} {\bibfnamefont {S.}~\bibnamefont
  {Lepri}}, \bibinfo {author} {\bibfnamefont {R.}~\bibnamefont {Livi}}, \ and\
  \bibinfo {author} {\bibfnamefont {A.}~\bibnamefont {Politi}},\ }\href
  {\doibase http://doi.org/10.1016/S0370-1573(02)00558-6} {\bibfield  {journal}
  {\bibinfo  {journal} {Phys. Rep.}\ }\textbf {\bibinfo {volume} {377}},\
  \bibinfo {pages} {1 } (\bibinfo {year} {2003})}\BibitemShut {NoStop}%
\bibitem [{\citenamefont {Lepri}(2016)}]{Lepri2016}%
  \BibitemOpen
  \bibinfo {editor} {\bibfnamefont {S.}~\bibnamefont {Lepri}},\ ed.,\
  \href@noop {} {\emph {\bibinfo {title} {{Thermal transport in low dimensions:
  from statistical physics to nanoscale heat transfer}}}},\ \bibinfo {series}
  {Lecture Notes in Physics}, Vol.\ \bibinfo {volume} {921}\ (\bibinfo
  {publisher} {Springer},\ \bibinfo {address} {Berlin},\ \bibinfo {year}
  {2016})\BibitemShut {NoStop}%
\bibitem [{\citenamefont {Lepri}\ \emph {et~al.}(1997)\citenamefont {Lepri},
  \citenamefont {Livi},\ and\ \citenamefont {Politi}}]{Lepri19971896}%
  \BibitemOpen
  \bibfield  {author} {\bibinfo {author} {\bibfnamefont {S.}~\bibnamefont
  {Lepri}}, \bibinfo {author} {\bibfnamefont {R.}~\bibnamefont {Livi}}, \ and\
  \bibinfo {author} {\bibfnamefont {A.}~\bibnamefont {Politi}},\ }\href
  {\doibase 10.1103/PhysRevLett.78.1896} {\bibfield  {journal} {\bibinfo
  {journal} {Phys. Rev. Lett.}\ }\textbf {\bibinfo {volume} {78}},\ \bibinfo
  {pages} {1896} (\bibinfo {year} {1997})}\BibitemShut {NoStop}%
\bibitem [{\citenamefont {Prosen}\ and\ \citenamefont
  {Campbell}(2000)}]{Prosen20002857}%
  \BibitemOpen
  \bibfield  {author} {\bibinfo {author} {\bibfnamefont {T.}~\bibnamefont
  {Prosen}}\ and\ \bibinfo {author} {\bibfnamefont {D.~K.}\ \bibnamefont
  {Campbell}},\ }\href {\doibase 10.1103/PhysRevLett.84.2857} {\bibfield
  {journal} {\bibinfo  {journal} {Phys. Rev. Lett.}\ }\textbf {\bibinfo
  {volume} {84}},\ \bibinfo {pages} {2857} (\bibinfo {year}
  {2000})}\BibitemShut {NoStop}%
\bibitem [{\citenamefont {Prosen}\ and\ \citenamefont
  {Campbell}(2005)}]{Prosen2005}%
  \BibitemOpen
  \bibfield  {author} {\bibinfo {author} {\bibfnamefont {T.}~\bibnamefont
  {Prosen}}\ and\ \bibinfo {author} {\bibfnamefont {D.~K.}\ \bibnamefont
  {Campbell}},\ }\href {\doibase 10.1063/1.1868532} {\bibfield  {journal}
  {\bibinfo  {journal} {Chaos}\ }\textbf {\bibinfo {volume} {15}},\ \bibinfo
  {pages} {015117} (\bibinfo {year} {2005})}\BibitemShut {NoStop}%
\bibitem [{\citenamefont {Narayan}\ and\ \citenamefont
  {Ramaswamy}(2002)}]{Narayan2002200601}%
  \BibitemOpen
  \bibfield  {author} {\bibinfo {author} {\bibfnamefont {O.}~\bibnamefont
  {Narayan}}\ and\ \bibinfo {author} {\bibfnamefont {S.}~\bibnamefont
  {Ramaswamy}},\ }\href {\doibase 10.1103/PhysRevLett.89.200601} {\bibfield
  {journal} {\bibinfo  {journal} {Phys. Rev. Lett.}\ }\textbf {\bibinfo
  {volume} {89}},\ \bibinfo {pages} {200601} (\bibinfo {year}
  {2002})}\BibitemShut {NoStop}%
\bibitem [{\citenamefont {Mai}\ and\ \citenamefont
  {Narayan}(2006)}]{Mai2006061202}%
  \BibitemOpen
  \bibfield  {author} {\bibinfo {author} {\bibfnamefont {T.}~\bibnamefont
  {Mai}}\ and\ \bibinfo {author} {\bibfnamefont {O.}~\bibnamefont {Narayan}},\
  }\href {\doibase 10.1103/PhysRevE.73.061202} {\bibfield  {journal} {\bibinfo
  {journal} {Phys. Rev. E}\ }\textbf {\bibinfo {volume} {73}},\ \bibinfo
  {pages} {061202} (\bibinfo {year} {2006})}\BibitemShut {NoStop}%
\bibitem [{\citenamefont {Mai}\ \emph {et~al.}(2007)\citenamefont {Mai},
  \citenamefont {Dhar},\ and\ \citenamefont {Narayan}}]{Mai2007184301}%
  \BibitemOpen
  \bibfield  {author} {\bibinfo {author} {\bibfnamefont {T.}~\bibnamefont
  {Mai}}, \bibinfo {author} {\bibfnamefont {A.}~\bibnamefont {Dhar}}, \ and\
  \bibinfo {author} {\bibfnamefont {O.}~\bibnamefont {Narayan}},\ }\href
  {\doibase 10.1103/PhysRevLett.98.184301} {\bibfield  {journal} {\bibinfo
  {journal} {Phys. Rev. Lett.}\ }\textbf {\bibinfo {volume} {98}},\ \bibinfo
  {pages} {184301} (\bibinfo {year} {2007})}\BibitemShut {NoStop}%
\bibitem [{\citenamefont {Lee-Dadswell}\ \emph {et~al.}(2005)\citenamefont
  {Lee-Dadswell}, \citenamefont {Nickel},\ and\ \citenamefont
  {Gray}}]{Lee2005031202}%
  \BibitemOpen
  \bibfield  {author} {\bibinfo {author} {\bibfnamefont {G.~R.}\ \bibnamefont
  {Lee-Dadswell}}, \bibinfo {author} {\bibfnamefont {B.~G.}\ \bibnamefont
  {Nickel}}, \ and\ \bibinfo {author} {\bibfnamefont {C.~G.}\ \bibnamefont
  {Gray}},\ }\href {\doibase 10.1103/PhysRevE.72.031202} {\bibfield  {journal}
  {\bibinfo  {journal} {Phys. Rev. E}\ }\textbf {\bibinfo {volume} {72}},\
  \bibinfo {pages} {031202} (\bibinfo {year} {2005})}\BibitemShut {NoStop}%
\bibitem [{\citenamefont {Lee-Dadswell}\ \emph {et~al.}(2008)\citenamefont
  {Lee-Dadswell}, \citenamefont {Nickel},\ and\ \citenamefont
  {Gray}}]{Lee-Dadswell2008}%
  \BibitemOpen
  \bibfield  {author} {\bibinfo {author} {\bibfnamefont {G.~R.}\ \bibnamefont
  {Lee-Dadswell}}, \bibinfo {author} {\bibfnamefont {B.~G.}\ \bibnamefont
  {Nickel}}, \ and\ \bibinfo {author} {\bibfnamefont {C.~G.}\ \bibnamefont
  {Gray}},\ }\href {\doibase 10.1007/s10955-008-9551-x} {\bibfield  {journal}
  {\bibinfo  {journal} {J. Stat. Phys.}\ }\textbf {\bibinfo {volume} {132}},\
  \bibinfo {pages} {1} (\bibinfo {year} {2008})}\BibitemShut {NoStop}%
\bibitem [{\citenamefont {Delfini}\ \emph {et~al.}(2006)\citenamefont
  {Delfini}, \citenamefont {Lepri}, \citenamefont {Livi},\ and\ \citenamefont
  {Politi}}]{Delfini2006060201}%
  \BibitemOpen
  \bibfield  {author} {\bibinfo {author} {\bibfnamefont {L.}~\bibnamefont
  {Delfini}}, \bibinfo {author} {\bibfnamefont {S.}~\bibnamefont {Lepri}},
  \bibinfo {author} {\bibfnamefont {R.}~\bibnamefont {Livi}}, \ and\ \bibinfo
  {author} {\bibfnamefont {A.}~\bibnamefont {Politi}},\ }\href {\doibase
  10.1103/PhysRevE.73.060201} {\bibfield  {journal} {\bibinfo  {journal} {Phys.
  Rev. E}\ }\textbf {\bibinfo {volume} {73}},\ \bibinfo {pages} {060201}
  (\bibinfo {year} {2006})}\BibitemShut {NoStop}%
\bibitem [{\citenamefont {Delfini}\ \emph {et~al.}(2007)\citenamefont
  {Delfini}, \citenamefont {Lepri}, \citenamefont {Livi},\ and\ \citenamefont
  {Politi}}]{DelfiniP02007}%
  \BibitemOpen
  \bibfield  {author} {\bibinfo {author} {\bibfnamefont {L.}~\bibnamefont
  {Delfini}}, \bibinfo {author} {\bibfnamefont {S.}~\bibnamefont {Lepri}},
  \bibinfo {author} {\bibfnamefont {R.}~\bibnamefont {Livi}}, \ and\ \bibinfo
  {author} {\bibfnamefont {A.}~\bibnamefont {Politi}},\ }\href
  {http://stacks.iop.org/1742-5468/2007/i=02/a=P02007} {\bibfield  {journal}
  {\bibinfo  {journal} {J. Stat. Mech.}\ }\textbf {\bibinfo {volume} {2007}},\
  \bibinfo {pages} {P02007} (\bibinfo {year} {2007})}\BibitemShut {NoStop}%
\bibitem [{\citenamefont {Wang}\ and\ \citenamefont
  {Li}(2004{\natexlab{a}})}]{Wang2004074302}%
  \BibitemOpen
  \bibfield  {author} {\bibinfo {author} {\bibfnamefont {J.-S.}\ \bibnamefont
  {Wang}}\ and\ \bibinfo {author} {\bibfnamefont {B.}~\bibnamefont {Li}},\
  }\href {\doibase 10.1103/PhysRevLett.92.074302} {\bibfield  {journal}
  {\bibinfo  {journal} {Phys. Rev. Lett.}\ }\textbf {\bibinfo {volume} {92}},\
  \bibinfo {pages} {074302} (\bibinfo {year} {2004}{\natexlab{a}})}\BibitemShut
  {NoStop}%
\bibitem [{\citenamefont {Wang}\ and\ \citenamefont
  {Li}(2004{\natexlab{b}})}]{Wang2004021204}%
  \BibitemOpen
  \bibfield  {author} {\bibinfo {author} {\bibfnamefont {J.-S.}\ \bibnamefont
  {Wang}}\ and\ \bibinfo {author} {\bibfnamefont {B.}~\bibnamefont {Li}},\
  }\href {\doibase 10.1103/PhysRevE.70.021204} {\bibfield  {journal} {\bibinfo
  {journal} {Phys. Rev. E}\ }\textbf {\bibinfo {volume} {70}},\ \bibinfo
  {pages} {021204} (\bibinfo {year} {2004}{\natexlab{b}})}\BibitemShut
  {NoStop}%
\bibitem [{\citenamefont {Hu}\ \emph {et~al.}(1998)\citenamefont {Hu},
  \citenamefont {Li},\ and\ \citenamefont {Zhao}}]{Hu19982992}%
  \BibitemOpen
  \bibfield  {author} {\bibinfo {author} {\bibfnamefont {B.}~\bibnamefont
  {Hu}}, \bibinfo {author} {\bibfnamefont {B.}~\bibnamefont {Li}}, \ and\
  \bibinfo {author} {\bibfnamefont {H.}~\bibnamefont {Zhao}},\ }\href {\doibase
  10.1103/PhysRevE.57.2992} {\bibfield  {journal} {\bibinfo  {journal} {Phys.
  Rev. E}\ }\textbf {\bibinfo {volume} {57}},\ \bibinfo {pages} {2992}
  (\bibinfo {year} {1998})}\BibitemShut {NoStop}%
\bibitem [{\citenamefont {Hu}\ \emph {et~al.}(2000)\citenamefont {Hu},
  \citenamefont {Li},\ and\ \citenamefont {Zhao}}]{Hu20003828}%
  \BibitemOpen
  \bibfield  {author} {\bibinfo {author} {\bibfnamefont {B.}~\bibnamefont
  {Hu}}, \bibinfo {author} {\bibfnamefont {B.}~\bibnamefont {Li}}, \ and\
  \bibinfo {author} {\bibfnamefont {H.}~\bibnamefont {Zhao}},\ }\href {\doibase
  10.1103/PhysRevE.61.3828} {\bibfield  {journal} {\bibinfo  {journal} {Phys.
  Rev. E}\ }\textbf {\bibinfo {volume} {61}},\ \bibinfo {pages} {3828}
  (\bibinfo {year} {2000})}\BibitemShut {NoStop}%
\bibitem [{\citenamefont {Zhong}\ \emph {et~al.}(2012)\citenamefont {Zhong},
  \citenamefont {Zhang}, \citenamefont {Wang},\ and\ \citenamefont
  {Zhao}}]{Zhong2012060102}%
  \BibitemOpen
  \bibfield  {author} {\bibinfo {author} {\bibfnamefont {Y.}~\bibnamefont
  {Zhong}}, \bibinfo {author} {\bibfnamefont {Y.}~\bibnamefont {Zhang}},
  \bibinfo {author} {\bibfnamefont {J.}~\bibnamefont {Wang}}, \ and\ \bibinfo
  {author} {\bibfnamefont {H.}~\bibnamefont {Zhao}},\ }\href {\doibase
  10.1103/PhysRevE.85.060102} {\bibfield  {journal} {\bibinfo  {journal} {Phys.
  Rev. E}\ }\textbf {\bibinfo {volume} {85}},\ \bibinfo {pages} {060102}
  (\bibinfo {year} {2012})}\BibitemShut {NoStop}%
\bibitem [{\citenamefont {{Zhang}}\ \emph {et~al.}(2013)\citenamefont
  {{Zhang}}, \citenamefont {{Chen}}, \citenamefont {{Wang}},\ and\
  \citenamefont {{Zhao}}}]{Zhang2013}%
  \BibitemOpen
  \bibfield  {author} {\bibinfo {author} {\bibfnamefont {Y.}~\bibnamefont
  {{Zhang}}}, \bibinfo {author} {\bibfnamefont {S.}~\bibnamefont {{Chen}}},
  \bibinfo {author} {\bibfnamefont {J.}~\bibnamefont {{Wang}}}, \ and\ \bibinfo
  {author} {\bibfnamefont {H.}~\bibnamefont {{Zhao}}},\ }\href@noop {}
  {\bibfield  {journal} {\bibinfo  {journal} {ArXiv e-prints}\ } (\bibinfo
  {year} {2013})},\ \Eprint {http://arxiv.org/abs/1301.2838} {arXiv:1301.2838
  [cond-mat.stat-mech]} \BibitemShut {NoStop}%
\bibitem [{\citenamefont {Wang}\ \emph {et~al.}(2013)\citenamefont {Wang},
  \citenamefont {Hu},\ and\ \citenamefont {Li}}]{Wang2013052112}%
  \BibitemOpen
  \bibfield  {author} {\bibinfo {author} {\bibfnamefont {L.}~\bibnamefont
  {Wang}}, \bibinfo {author} {\bibfnamefont {B.}~\bibnamefont {Hu}}, \ and\
  \bibinfo {author} {\bibfnamefont {B.}~\bibnamefont {Li}},\ }\href {\doibase
  10.1103/PhysRevE.88.052112} {\bibfield  {journal} {\bibinfo  {journal} {Phys.
  Rev. E}\ }\textbf {\bibinfo {volume} {88}},\ \bibinfo {pages} {052112}
  (\bibinfo {year} {2013})}\BibitemShut {NoStop}%
\bibitem [{\citenamefont {Savin}\ and\ \citenamefont
  {Kosevich}(2014)}]{Savin2014032102}%
  \BibitemOpen
  \bibfield  {author} {\bibinfo {author} {\bibfnamefont {A.~V.}\ \bibnamefont
  {Savin}}\ and\ \bibinfo {author} {\bibfnamefont {Y.~A.}\ \bibnamefont
  {Kosevich}},\ }\href {\doibase 10.1103/PhysRevE.89.032102} {\bibfield
  {journal} {\bibinfo  {journal} {Phys. Rev. E}\ }\textbf {\bibinfo {volume}
  {89}},\ \bibinfo {pages} {032102} (\bibinfo {year} {2014})}\BibitemShut
  {NoStop}%
\bibitem [{\citenamefont {Das}\ \emph {et~al.}(2014{\natexlab{a}})\citenamefont
  {Das}, \citenamefont {Dhar},\ and\ \citenamefont {Narayan}}]{Das2014204}%
  \BibitemOpen
  \bibfield  {author} {\bibinfo {author} {\bibfnamefont {S.~G.}\ \bibnamefont
  {Das}}, \bibinfo {author} {\bibfnamefont {A.}~\bibnamefont {Dhar}}, \ and\
  \bibinfo {author} {\bibfnamefont {O.}~\bibnamefont {Narayan}},\ }\href
  {\doibase 10.1007/s10955-013-0871-0} {\bibfield  {journal} {\bibinfo
  {journal} {J. Stat. Phys.}\ }\textbf {\bibinfo {volume} {154}},\ \bibinfo
  {pages} {204} (\bibinfo {year} {2014}{\natexlab{a}})}\BibitemShut {NoStop}%
\bibitem [{\citenamefont {Chang}\ \emph {et~al.}(2008)\citenamefont {Chang},
  \citenamefont {Okawa}, \citenamefont {Garcia}, \citenamefont {Majumdar},\
  and\ \citenamefont {Zettl}}]{Chang2008075903}%
  \BibitemOpen
  \bibfield  {author} {\bibinfo {author} {\bibfnamefont {C.~W.}\ \bibnamefont
  {Chang}}, \bibinfo {author} {\bibfnamefont {D.}~\bibnamefont {Okawa}},
  \bibinfo {author} {\bibfnamefont {H.}~\bibnamefont {Garcia}}, \bibinfo
  {author} {\bibfnamefont {A.}~\bibnamefont {Majumdar}}, \ and\ \bibinfo
  {author} {\bibfnamefont {A.}~\bibnamefont {Zettl}},\ }\href {\doibase
  10.1103/PhysRevLett.101.075903} {\bibfield  {journal} {\bibinfo  {journal}
  {Phys. Rev. Lett.}\ }\textbf {\bibinfo {volume} {101}},\ \bibinfo {pages}
  {075903} (\bibinfo {year} {2008})}\BibitemShut {NoStop}%
\bibitem [{\citenamefont {Lee}\ \emph {et~al.}(2017)\citenamefont {Lee},
  \citenamefont {Wu}, \citenamefont {Lou}, \citenamefont {Lee},\ and\
  \citenamefont {Chang}}]{Lee2017135901}%
  \BibitemOpen
  \bibfield  {author} {\bibinfo {author} {\bibfnamefont {V.}~\bibnamefont
  {Lee}}, \bibinfo {author} {\bibfnamefont {C.-H.}\ \bibnamefont {Wu}},
  \bibinfo {author} {\bibfnamefont {Z.-X.}\ \bibnamefont {Lou}}, \bibinfo
  {author} {\bibfnamefont {W.-L.}\ \bibnamefont {Lee}}, \ and\ \bibinfo
  {author} {\bibfnamefont {C.-W.}\ \bibnamefont {Chang}},\ }\href {\doibase
  10.1103/PhysRevLett.118.135901} {\bibfield  {journal} {\bibinfo  {journal}
  {Phys. Rev. Lett.}\ }\textbf {\bibinfo {volume} {118}},\ \bibinfo {pages}
  {135901} (\bibinfo {year} {2017})}\BibitemShut {NoStop}%
\bibitem [{\citenamefont {{Xu}}\ \emph {et~al.}(2014)\citenamefont {{Xu}},
  \citenamefont {{Pereira}}, \citenamefont {{Wang}}, \citenamefont {{Wu}},
  \citenamefont {{Zhang}}, \citenamefont {{Zhao}}, \citenamefont {{Bae}},
  \citenamefont {{Tinh Bui}}, \citenamefont {{Xie}}, \citenamefont {{Thong}},
  \citenamefont {{Hong}}, \citenamefont {{Loh}}, \citenamefont {{Donadio}},
  \citenamefont {{Li}},\ and\ \citenamefont {{{\"O}zyilmaz}}}]{Xu2014}%
  \BibitemOpen
  \bibfield  {author} {\bibinfo {author} {\bibfnamefont {X.}~\bibnamefont
  {{Xu}}}, \bibinfo {author} {\bibfnamefont {L.~F.~C.}\ \bibnamefont
  {{Pereira}}}, \bibinfo {author} {\bibfnamefont {Y.}~\bibnamefont {{Wang}}},
  \bibinfo {author} {\bibfnamefont {J.}~\bibnamefont {{Wu}}}, \bibinfo {author}
  {\bibfnamefont {K.}~\bibnamefont {{Zhang}}}, \bibinfo {author} {\bibfnamefont
  {X.}~\bibnamefont {{Zhao}}}, \bibinfo {author} {\bibfnamefont
  {S.}~\bibnamefont {{Bae}}}, \bibinfo {author} {\bibfnamefont
  {C.}~\bibnamefont {{Tinh Bui}}}, \bibinfo {author} {\bibfnamefont
  {R.}~\bibnamefont {{Xie}}}, \bibinfo {author} {\bibfnamefont {J.~T.~L.}\
  \bibnamefont {{Thong}}}, \bibinfo {author} {\bibfnamefont {B.~H.}\
  \bibnamefont {{Hong}}}, \bibinfo {author} {\bibfnamefont {K.~P.}\
  \bibnamefont {{Loh}}}, \bibinfo {author} {\bibfnamefont {D.}~\bibnamefont
  {{Donadio}}}, \bibinfo {author} {\bibfnamefont {B.}~\bibnamefont {{Li}}}, \
  and\ \bibinfo {author} {\bibfnamefont {B.}~\bibnamefont {{{\"O}zyilmaz}}},\
  }\href {\doibase 10.1038/ncomms4689} {\bibfield  {journal} {\bibinfo
  {journal} {Nat. Commun.}\ }\textbf {\bibinfo {volume} {5}},\ \bibinfo {eid}
  {3689} (\bibinfo {year} {2014})}\BibitemShut {NoStop}%
\bibitem [{\citenamefont {Denisov}\ \emph {et~al.}(2003)\citenamefont
  {Denisov}, \citenamefont {Klafter},\ and\ \citenamefont
  {Urbakh}}]{Denisov2003194301}%
  \BibitemOpen
  \bibfield  {author} {\bibinfo {author} {\bibfnamefont {S.}~\bibnamefont
  {Denisov}}, \bibinfo {author} {\bibfnamefont {J.}~\bibnamefont {Klafter}}, \
  and\ \bibinfo {author} {\bibfnamefont {M.}~\bibnamefont {Urbakh}},\ }\href
  {\doibase 10.1103/PhysRevLett.91.194301} {\bibfield  {journal} {\bibinfo
  {journal} {Phys. Rev. Lett.}\ }\textbf {\bibinfo {volume} {91}},\ \bibinfo
  {pages} {194301} (\bibinfo {year} {2003})}\BibitemShut {NoStop}%
\bibitem [{\citenamefont {Cipriani}\ \emph {et~al.}(2005)\citenamefont
  {Cipriani}, \citenamefont {Denisov},\ and\ \citenamefont
  {Politi}}]{Cipriani2005}%
  \BibitemOpen
  \bibfield  {author} {\bibinfo {author} {\bibfnamefont {P.}~\bibnamefont
  {Cipriani}}, \bibinfo {author} {\bibfnamefont {S.}~\bibnamefont {Denisov}}, \
  and\ \bibinfo {author} {\bibfnamefont {A.}~\bibnamefont {Politi}},\ }\href
  {\doibase 10.1103/PhysRevLett.94.244301} {\bibfield  {journal} {\bibinfo
  {journal} {Phys. Rev. Lett.}\ }\textbf {\bibinfo {volume} {94}},\ \bibinfo
  {pages} {244301} (\bibinfo {year} {2005})}\BibitemShut {NoStop}%
\bibitem [{\citenamefont {Zaburdaev}\ \emph {et~al.}(2011)\citenamefont
  {Zaburdaev}, \citenamefont {Denisov},\ and\ \citenamefont
  {H\"anggi}}]{Zaburdaev2011}%
  \BibitemOpen
  \bibfield  {author} {\bibinfo {author} {\bibfnamefont {V.}~\bibnamefont
  {Zaburdaev}}, \bibinfo {author} {\bibfnamefont {S.}~\bibnamefont {Denisov}},
  \ and\ \bibinfo {author} {\bibfnamefont {P.}~\bibnamefont {H\"anggi}},\
  }\href {\doibase 10.1103/PhysRevLett.106.180601} {\bibfield  {journal}
  {\bibinfo  {journal} {Phys. Rev. Lett.}\ }\textbf {\bibinfo {volume} {106}},\
  \bibinfo {pages} {180601} (\bibinfo {year} {2011})}\BibitemShut {NoStop}%
\bibitem [{\citenamefont {Denisov}\ \emph {et~al.}(2012)\citenamefont
  {Denisov}, \citenamefont {Zaburdaev},\ and\ \citenamefont
  {H\"anggi}}]{Denisov2012031148}%
  \BibitemOpen
  \bibfield  {author} {\bibinfo {author} {\bibfnamefont {S.}~\bibnamefont
  {Denisov}}, \bibinfo {author} {\bibfnamefont {V.}~\bibnamefont {Zaburdaev}},
  \ and\ \bibinfo {author} {\bibfnamefont {P.}~\bibnamefont {H\"anggi}},\
  }\href {\doibase 10.1103/PhysRevE.85.031148} {\bibfield  {journal} {\bibinfo
  {journal} {Phys. Rev. E}\ }\textbf {\bibinfo {volume} {85}},\ \bibinfo
  {pages} {031148} (\bibinfo {year} {2012})}\BibitemShut {NoStop}%
\bibitem [{\citenamefont {Zaburdaev}\ \emph {et~al.}(2013)\citenamefont
  {Zaburdaev}, \citenamefont {Denisov},\ and\ \citenamefont
  {H\"anggi}}]{Zaburdaev2013}%
  \BibitemOpen
  \bibfield  {author} {\bibinfo {author} {\bibfnamefont {V.}~\bibnamefont
  {Zaburdaev}}, \bibinfo {author} {\bibfnamefont {S.}~\bibnamefont {Denisov}},
  \ and\ \bibinfo {author} {\bibfnamefont {P.}~\bibnamefont {H\"anggi}},\
  }\href {\doibase 10.1103/PhysRevLett.110.170604} {\bibfield  {journal}
  {\bibinfo  {journal} {Phys. Rev. Lett.}\ }\textbf {\bibinfo {volume} {110}},\
  \bibinfo {pages} {170604} (\bibinfo {year} {2013})}\BibitemShut {NoStop}%
\bibitem [{\citenamefont {Zaburdaev}\ \emph {et~al.}(2015)\citenamefont
  {Zaburdaev}, \citenamefont {Denisov},\ and\ \citenamefont
  {Klafter}}]{Zaburdaev2015}%
  \BibitemOpen
  \bibfield  {author} {\bibinfo {author} {\bibfnamefont {V.}~\bibnamefont
  {Zaburdaev}}, \bibinfo {author} {\bibfnamefont {S.}~\bibnamefont {Denisov}},
  \ and\ \bibinfo {author} {\bibfnamefont {J.}~\bibnamefont {Klafter}},\ }\href
  {\doibase 10.1103/RevModPhys.87.483} {\bibfield  {journal} {\bibinfo
  {journal} {Rev. Mod. Phys.}\ }\textbf {\bibinfo {volume} {87}},\ \bibinfo
  {pages} {483} (\bibinfo {year} {2015})}\BibitemShut {NoStop}%
\bibitem [{\citenamefont {Liu}\ \emph {et~al.}(2014{\natexlab{a}})\citenamefont
  {Liu}, \citenamefont {H\"anggi}, \citenamefont {Li}, \citenamefont {Ren},\
  and\ \citenamefont {Li}}]{Liu2014}%
  \BibitemOpen
  \bibfield  {author} {\bibinfo {author} {\bibfnamefont {S.}~\bibnamefont
  {Liu}}, \bibinfo {author} {\bibfnamefont {P.}~\bibnamefont {H\"anggi}},
  \bibinfo {author} {\bibfnamefont {N.}~\bibnamefont {Li}}, \bibinfo {author}
  {\bibfnamefont {J.}~\bibnamefont {Ren}}, \ and\ \bibinfo {author}
  {\bibfnamefont {B.}~\bibnamefont {Li}},\ }\href {\doibase
  10.1103/PhysRevLett.112.040601} {\bibfield  {journal} {\bibinfo  {journal}
  {Phys. Rev. Lett.}\ }\textbf {\bibinfo {volume} {112}},\ \bibinfo {pages}
  {040601} (\bibinfo {year} {2014}{\natexlab{a}})}\BibitemShut {NoStop}%
\bibitem [{\citenamefont {Dhar}\ \emph {et~al.}(2013)\citenamefont {Dhar},
  \citenamefont {Saito},\ and\ \citenamefont {Derrida}}]{Dhar2013010103}%
  \BibitemOpen
  \bibfield  {author} {\bibinfo {author} {\bibfnamefont {A.}~\bibnamefont
  {Dhar}}, \bibinfo {author} {\bibfnamefont {K.}~\bibnamefont {Saito}}, \ and\
  \bibinfo {author} {\bibfnamefont {B.}~\bibnamefont {Derrida}},\ }\href
  {\doibase 10.1103/PhysRevE.87.010103} {\bibfield  {journal} {\bibinfo
  {journal} {Phys. Rev. E}\ }\textbf {\bibinfo {volume} {87}},\ \bibinfo
  {pages} {010103} (\bibinfo {year} {2013})}\BibitemShut {NoStop}%
\bibitem [{\citenamefont {Lepri}\ and\ \citenamefont
  {Politi}(2011)}]{Lepri2011030107}%
  \BibitemOpen
  \bibfield  {author} {\bibinfo {author} {\bibfnamefont {S.}~\bibnamefont
  {Lepri}}\ and\ \bibinfo {author} {\bibfnamefont {A.}~\bibnamefont {Politi}},\
  }\href {\doibase 10.1103/PhysRevE.83.030107} {\bibfield  {journal} {\bibinfo
  {journal} {Phys. Rev. E}\ }\textbf {\bibinfo {volume} {83}},\ \bibinfo
  {pages} {030107} (\bibinfo {year} {2011})}\BibitemShut {NoStop}%
\bibitem [{\citenamefont {Mendl}\ and\ \citenamefont
  {Spohn}(2013)}]{Mendl2013}%
  \BibitemOpen
  \bibfield  {author} {\bibinfo {author} {\bibfnamefont {C.~B.}\ \bibnamefont
  {Mendl}}\ and\ \bibinfo {author} {\bibfnamefont {H.}~\bibnamefont {Spohn}},\
  }\href {\doibase 10.1103/PhysRevLett.111.230601} {\bibfield  {journal}
  {\bibinfo  {journal} {Phys. Rev. Lett.}\ }\textbf {\bibinfo {volume} {111}},\
  \bibinfo {pages} {230601} (\bibinfo {year} {2013})}\BibitemShut {NoStop}%
\bibitem [{\citenamefont {Mendl}\ and\ \citenamefont
  {Spohn}(2014)}]{Mendl2014012147}%
  \BibitemOpen
  \bibfield  {author} {\bibinfo {author} {\bibfnamefont {C.~B.}\ \bibnamefont
  {Mendl}}\ and\ \bibinfo {author} {\bibfnamefont {H.}~\bibnamefont {Spohn}},\
  }\href {\doibase 10.1103/PhysRevE.90.012147} {\bibfield  {journal} {\bibinfo
  {journal} {Phys. Rev. E}\ }\textbf {\bibinfo {volume} {90}},\ \bibinfo
  {pages} {012147} (\bibinfo {year} {2014})}\BibitemShut {NoStop}%
\bibitem [{\citenamefont {Spohn}(2014)}]{Spohn2014}%
  \BibitemOpen
  \bibfield  {author} {\bibinfo {author} {\bibfnamefont {H.}~\bibnamefont
  {Spohn}},\ }\href {\doibase 10.1007/s10955-014-0933-y} {\bibfield  {journal}
  {\bibinfo  {journal} {J. Stat. Phys.}\ }\textbf {\bibinfo {volume} {154}},\
  \bibinfo {pages} {1191} (\bibinfo {year} {2014})}\BibitemShut {NoStop}%
\bibitem [{\citenamefont {Das}\ \emph {et~al.}(2014{\natexlab{b}})\citenamefont
  {Das}, \citenamefont {Dhar}, \citenamefont {Saito}, \citenamefont {Mendl},\
  and\ \citenamefont {Spohn}}]{Das2014}%
  \BibitemOpen
  \bibfield  {author} {\bibinfo {author} {\bibfnamefont {S.~G.}\ \bibnamefont
  {Das}}, \bibinfo {author} {\bibfnamefont {A.}~\bibnamefont {Dhar}}, \bibinfo
  {author} {\bibfnamefont {K.}~\bibnamefont {Saito}}, \bibinfo {author}
  {\bibfnamefont {C.~B.}\ \bibnamefont {Mendl}}, \ and\ \bibinfo {author}
  {\bibfnamefont {H.}~\bibnamefont {Spohn}},\ }\href {\doibase
  10.1103/PhysRevE.90.012124} {\bibfield  {journal} {\bibinfo  {journal} {Phys.
  Rev. E}\ }\textbf {\bibinfo {volume} {90}},\ \bibinfo {pages} {012124}
  (\bibinfo {year} {2014}{\natexlab{b}})}\BibitemShut {NoStop}%
\bibitem [{\citenamefont {van Beijeren}(2012)}]{van2012}%
  \BibitemOpen
  \bibfield  {author} {\bibinfo {author} {\bibfnamefont {H.}~\bibnamefont {van
  Beijeren}},\ }\href {\doibase 10.1103/PhysRevLett.108.180601} {\bibfield
  {journal} {\bibinfo  {journal} {Phys. Rev. Lett.}\ }\textbf {\bibinfo
  {volume} {108}},\ \bibinfo {pages} {180601} (\bibinfo {year}
  {2012})}\BibitemShut {NoStop}%
\bibitem [{\citenamefont {Flach}\ and\ \citenamefont
  {Gorbach}(2008)}]{Flach20081}%
  \BibitemOpen
  \bibfield  {author} {\bibinfo {author} {\bibfnamefont {S.}~\bibnamefont
  {Flach}}\ and\ \bibinfo {author} {\bibfnamefont {A.~V.}\ \bibnamefont
  {Gorbach}},\ }\href {\doibase
  http://dx.doi.org/10.1016/j.physrep.2008.05.002} {\bibfield  {journal}
  {\bibinfo  {journal} {Phys. Rep.}\ }\textbf {\bibinfo {volume} {467}},\
  \bibinfo {pages} {1 } (\bibinfo {year} {2008})}\BibitemShut {NoStop}%
\bibitem [{\citenamefont {Szeftel}\ \emph {et~al.}(2003)\citenamefont
  {Szeftel}, \citenamefont {Huang},\ and\ \citenamefont
  {Konotop}}]{SZEFTEL2003215}%
  \BibitemOpen
  \bibfield  {author} {\bibinfo {author} {\bibfnamefont {J.}~\bibnamefont
  {Szeftel}}, \bibinfo {author} {\bibfnamefont {G.}~\bibnamefont {Huang}}, \
  and\ \bibinfo {author} {\bibfnamefont {V.}~\bibnamefont {Konotop}},\ }\href
  {\doibase https://doi.org/10.1016/S0167-2789(03)00116-7} {\bibfield
  {journal} {\bibinfo  {journal} {Physica D}\ }\textbf {\bibinfo {volume}
  {181}},\ \bibinfo {pages} {215 } (\bibinfo {year} {2003})}\BibitemShut
  {NoStop}%
\bibitem [{\citenamefont {Riseborough}(2012)}]{Riseborough2012011129}%
  \BibitemOpen
  \bibfield  {author} {\bibinfo {author} {\bibfnamefont {P.~S.}\ \bibnamefont
  {Riseborough}},\ }\href {\doibase 10.1103/PhysRevE.85.011129} {\bibfield
  {journal} {\bibinfo  {journal} {Phys. Rev. E}\ }\textbf {\bibinfo {volume}
  {85}},\ \bibinfo {pages} {011129} (\bibinfo {year} {2012})}\BibitemShut
  {NoStop}%
\bibitem [{\citenamefont {Tsironis}\ \emph {et~al.}(1999)\citenamefont
  {Tsironis}, \citenamefont {Bishop}, \citenamefont {Savin},\ and\
  \citenamefont {Zolotaryuk}}]{Tsironis19996610}%
  \BibitemOpen
  \bibfield  {author} {\bibinfo {author} {\bibfnamefont {G.~P.}\ \bibnamefont
  {Tsironis}}, \bibinfo {author} {\bibfnamefont {A.~R.}\ \bibnamefont
  {Bishop}}, \bibinfo {author} {\bibfnamefont {A.~V.}\ \bibnamefont {Savin}}, \
  and\ \bibinfo {author} {\bibfnamefont {A.~V.}\ \bibnamefont {Zolotaryuk}},\
  }\href {\doibase 10.1103/PhysRevE.60.6610} {\bibfield  {journal} {\bibinfo
  {journal} {Phys. Rev. E}\ }\textbf {\bibinfo {volume} {60}},\ \bibinfo
  {pages} {6610} (\bibinfo {year} {1999})}\BibitemShut {NoStop}%
\bibitem [{\citenamefont {Gendelman}\ and\ \citenamefont
  {Savin}(2000)}]{Gendelman20002381}%
  \BibitemOpen
  \bibfield  {author} {\bibinfo {author} {\bibfnamefont {O.~V.}\ \bibnamefont
  {Gendelman}}\ and\ \bibinfo {author} {\bibfnamefont {A.~V.}\ \bibnamefont
  {Savin}},\ }\href {\doibase 10.1103/PhysRevLett.84.2381} {\bibfield
  {journal} {\bibinfo  {journal} {Phys. Rev. Lett.}\ }\textbf {\bibinfo
  {volume} {84}},\ \bibinfo {pages} {2381} (\bibinfo {year}
  {2000})}\BibitemShut {NoStop}%
\bibitem [{\citenamefont {Giardin\`a}\ \emph {et~al.}(2000)\citenamefont
  {Giardin\`a}, \citenamefont {Livi}, \citenamefont {Politi},\ and\
  \citenamefont {Vassalli}}]{Giardina20002144}%
  \BibitemOpen
  \bibfield  {author} {\bibinfo {author} {\bibfnamefont {C.}~\bibnamefont
  {Giardin\`a}}, \bibinfo {author} {\bibfnamefont {R.}~\bibnamefont {Livi}},
  \bibinfo {author} {\bibfnamefont {A.}~\bibnamefont {Politi}}, \ and\ \bibinfo
  {author} {\bibfnamefont {M.}~\bibnamefont {Vassalli}},\ }\href {\doibase
  10.1103/PhysRevLett.84.2144} {\bibfield  {journal} {\bibinfo  {journal}
  {Phys. Rev. Lett.}\ }\textbf {\bibinfo {volume} {84}},\ \bibinfo {pages}
  {2144} (\bibinfo {year} {2000})}\BibitemShut {NoStop}%
\bibitem [{\citenamefont {Xiong}\ \emph {et~al.}(2012)\citenamefont {Xiong},
  \citenamefont {Wang}, \citenamefont {Zhang},\ and\ \citenamefont
  {Zhao}}]{Xiong2012020102}%
  \BibitemOpen
  \bibfield  {author} {\bibinfo {author} {\bibfnamefont {D.}~\bibnamefont
  {Xiong}}, \bibinfo {author} {\bibfnamefont {J.}~\bibnamefont {Wang}},
  \bibinfo {author} {\bibfnamefont {Y.}~\bibnamefont {Zhang}}, \ and\ \bibinfo
  {author} {\bibfnamefont {H.}~\bibnamefont {Zhao}},\ }\href {\doibase
  10.1103/PhysRevE.85.020102} {\bibfield  {journal} {\bibinfo  {journal} {Phys.
  Rev. E}\ }\textbf {\bibinfo {volume} {85}},\ \bibinfo {pages} {020102}
  (\bibinfo {year} {2012})}\BibitemShut {NoStop}%
\bibitem [{\citenamefont {Xiong}\ \emph {et~al.}(2014)\citenamefont {Xiong},
  \citenamefont {Zhang},\ and\ \citenamefont {Zhao}}]{Xiong2014022117}%
  \BibitemOpen
  \bibfield  {author} {\bibinfo {author} {\bibfnamefont {D.}~\bibnamefont
  {Xiong}}, \bibinfo {author} {\bibfnamefont {Y.}~\bibnamefont {Zhang}}, \ and\
  \bibinfo {author} {\bibfnamefont {H.}~\bibnamefont {Zhao}},\ }\href {\doibase
  10.1103/PhysRevE.90.022117} {\bibfield  {journal} {\bibinfo  {journal} {Phys.
  Rev. E}\ }\textbf {\bibinfo {volume} {90}},\ \bibinfo {pages} {022117}
  (\bibinfo {year} {2014})}\BibitemShut {NoStop}%
\bibitem [{\citenamefont {Fillipov}\ \emph {et~al.}(1998)\citenamefont
  {Fillipov}, \citenamefont {Hu}, \citenamefont {Li},\ and\ \citenamefont
  {Zeltser}}]{Fillipov1998}%
  \BibitemOpen
  \bibfield  {author} {\bibinfo {author} {\bibfnamefont {A.}~\bibnamefont
  {Fillipov}}, \bibinfo {author} {\bibfnamefont {B.}~\bibnamefont {Hu}},
  \bibinfo {author} {\bibfnamefont {B.}~\bibnamefont {Li}}, \ and\ \bibinfo
  {author} {\bibfnamefont {A.}~\bibnamefont {Zeltser}},\ }\href
  {http://stacks.iop.org/0305-4470/31/i=38/a=008} {\bibfield  {journal}
  {\bibinfo  {journal} {J. Phys. A}\ }\textbf {\bibinfo {volume} {31}},\
  \bibinfo {pages} {7719} (\bibinfo {year} {1998})}\BibitemShut {NoStop}%
\bibitem [{\citenamefont {Theodorakopoulos}\ and\ \citenamefont
  {Peyrard}(1999)}]{Theodorakopoulos19992293}%
  \BibitemOpen
  \bibfield  {author} {\bibinfo {author} {\bibfnamefont {N.}~\bibnamefont
  {Theodorakopoulos}}\ and\ \bibinfo {author} {\bibfnamefont {M.}~\bibnamefont
  {Peyrard}},\ }\href {\doibase 10.1103/PhysRevLett.83.2293} {\bibfield
  {journal} {\bibinfo  {journal} {Phys. Rev. Lett.}\ }\textbf {\bibinfo
  {volume} {83}},\ \bibinfo {pages} {2293} (\bibinfo {year}
  {1999})}\BibitemShut {NoStop}%
\bibitem [{\citenamefont {Zhang}\ \emph {et~al.}(2000)\citenamefont {Zhang},
  \citenamefont {Isbister},\ and\ \citenamefont {Evans}}]{Zhang20003541}%
  \BibitemOpen
  \bibfield  {author} {\bibinfo {author} {\bibfnamefont {F.}~\bibnamefont
  {Zhang}}, \bibinfo {author} {\bibfnamefont {D.~J.}\ \bibnamefont {Isbister}},
  \ and\ \bibinfo {author} {\bibfnamefont {D.~J.}\ \bibnamefont {Evans}},\
  }\href {\doibase 10.1103/PhysRevE.61.3541} {\bibfield  {journal} {\bibinfo
  {journal} {Phys. Rev. E}\ }\textbf {\bibinfo {volume} {61}},\ \bibinfo
  {pages} {3541} (\bibinfo {year} {2000})}\BibitemShut {NoStop}%
\bibitem [{\citenamefont {Aoki}\ and\ \citenamefont
  {Kusnezov}(2001)}]{Aoki20014029}%
  \BibitemOpen
  \bibfield  {author} {\bibinfo {author} {\bibfnamefont {K.}~\bibnamefont
  {Aoki}}\ and\ \bibinfo {author} {\bibfnamefont {D.}~\bibnamefont
  {Kusnezov}},\ }\href {\doibase 10.1103/PhysRevLett.86.4029} {\bibfield
  {journal} {\bibinfo  {journal} {Phys. Rev. Lett.}\ }\textbf {\bibinfo
  {volume} {86}},\ \bibinfo {pages} {4029} (\bibinfo {year}
  {2001})}\BibitemShut {NoStop}%
\bibitem [{\citenamefont {Li}\ \emph {et~al.}(2005)\citenamefont {Li},
  \citenamefont {Wang}, \citenamefont {Wang},\ and\ \citenamefont
  {Zhang}}]{Li2005}%
  \BibitemOpen
  \bibfield  {author} {\bibinfo {author} {\bibfnamefont {B.}~\bibnamefont
  {Li}}, \bibinfo {author} {\bibfnamefont {J.}~\bibnamefont {Wang}}, \bibinfo
  {author} {\bibfnamefont {L.}~\bibnamefont {Wang}}, \ and\ \bibinfo {author}
  {\bibfnamefont {G.}~\bibnamefont {Zhang}},\ }\href {\doibase
  10.1063/1.1832791} {\bibfield  {journal} {\bibinfo  {journal} {Chaos}\
  }\textbf {\bibinfo {volume} {15}},\ \bibinfo {pages} {015121} (\bibinfo
  {year} {2005})}\BibitemShut {NoStop}%
\bibitem [{\citenamefont {Zhao}\ \emph {et~al.}(2005)\citenamefont {Zhao},
  \citenamefont {Wen}, \citenamefont {Zhang},\ and\ \citenamefont
  {Zheng}}]{Zhao2005025507}%
  \BibitemOpen
  \bibfield  {author} {\bibinfo {author} {\bibfnamefont {H.}~\bibnamefont
  {Zhao}}, \bibinfo {author} {\bibfnamefont {Z.}~\bibnamefont {Wen}}, \bibinfo
  {author} {\bibfnamefont {Y.}~\bibnamefont {Zhang}}, \ and\ \bibinfo {author}
  {\bibfnamefont {D.}~\bibnamefont {Zheng}},\ }\href {\doibase
  10.1103/PhysRevLett.94.025507} {\bibfield  {journal} {\bibinfo  {journal}
  {Phys. Rev. Lett.}\ }\textbf {\bibinfo {volume} {94}},\ \bibinfo {pages}
  {025507} (\bibinfo {year} {2005})}\BibitemShut {NoStop}%
\bibitem [{\citenamefont {Zhao}(2006)}]{Zhao2006140602}%
  \BibitemOpen
  \bibfield  {author} {\bibinfo {author} {\bibfnamefont {H.}~\bibnamefont
  {Zhao}},\ }\href {\doibase 10.1103/PhysRevLett.96.140602} {\bibfield
  {journal} {\bibinfo  {journal} {Phys. Rev. Lett.}\ }\textbf {\bibinfo
  {volume} {96}},\ \bibinfo {pages} {140602} (\bibinfo {year}
  {2006})}\BibitemShut {NoStop}%
\bibitem [{\citenamefont {Li}\ \emph {et~al.}(2010)\citenamefont {Li},
  \citenamefont {Li},\ and\ \citenamefont {Flach}}]{Li2010054102}%
  \BibitemOpen
  \bibfield  {author} {\bibinfo {author} {\bibfnamefont {N.}~\bibnamefont
  {Li}}, \bibinfo {author} {\bibfnamefont {B.}~\bibnamefont {Li}}, \ and\
  \bibinfo {author} {\bibfnamefont {S.}~\bibnamefont {Flach}},\ }\href
  {\doibase 10.1103/PhysRevLett.105.054102} {\bibfield  {journal} {\bibinfo
  {journal} {Phys. Rev. Lett.}\ }\textbf {\bibinfo {volume} {105}},\ \bibinfo
  {pages} {054102} (\bibinfo {year} {2010})}\BibitemShut {NoStop}%
\bibitem [{\citenamefont {Li}\ and\ \citenamefont {Li}(2012)}]{Li2012AIP}%
  \BibitemOpen
  \bibfield  {author} {\bibinfo {author} {\bibfnamefont {N.}~\bibnamefont
  {Li}}\ and\ \bibinfo {author} {\bibfnamefont {B.}~\bibnamefont {Li}},\ }\href
  {\doibase 10.1063/1.4773459} {\bibfield  {journal} {\bibinfo  {journal} {AIP
  Adv.}\ }\textbf {\bibinfo {volume} {2}},\ \bibinfo {pages} {041408} (\bibinfo
  {year} {2012})}\BibitemShut {NoStop}%
\bibitem [{\citenamefont {Li}\ and\ \citenamefont {Li}(2013)}]{Li2013}%
  \BibitemOpen
  \bibfield  {author} {\bibinfo {author} {\bibfnamefont {N.}~\bibnamefont
  {Li}}\ and\ \bibinfo {author} {\bibfnamefont {B.}~\bibnamefont {Li}},\ }\href
  {\doibase 10.1103/PhysRevE.87.042125} {\bibfield  {journal} {\bibinfo
  {journal} {Phys. Rev. E}\ }\textbf {\bibinfo {volume} {87}},\ \bibinfo
  {pages} {042125} (\bibinfo {year} {2013})}\BibitemShut {NoStop}%
\bibitem [{\citenamefont {Liu}\ \emph {et~al.}(2014{\natexlab{b}})\citenamefont
  {Liu}, \citenamefont {Liu}, \citenamefont {H\"anggi}, \citenamefont {Wu},\
  and\ \citenamefont {Li}}]{Liu2014174304}%
  \BibitemOpen
  \bibfield  {author} {\bibinfo {author} {\bibfnamefont {S.}~\bibnamefont
  {Liu}}, \bibinfo {author} {\bibfnamefont {J.}~\bibnamefont {Liu}}, \bibinfo
  {author} {\bibfnamefont {P.}~\bibnamefont {H\"anggi}}, \bibinfo {author}
  {\bibfnamefont {C.}~\bibnamefont {Wu}}, \ and\ \bibinfo {author}
  {\bibfnamefont {B.}~\bibnamefont {Li}},\ }\href {\doibase
  10.1103/PhysRevB.90.174304} {\bibfield  {journal} {\bibinfo  {journal} {Phys.
  Rev. B}\ }\textbf {\bibinfo {volume} {90}},\ \bibinfo {pages} {174304}
  (\bibinfo {year} {2014}{\natexlab{b}})}\BibitemShut {NoStop}%
\bibitem [{\citenamefont {Liu}\ \emph {et~al.}(2015)\citenamefont {Liu},
  \citenamefont {Liu}, \citenamefont {Li}, \citenamefont {Li},\ and\
  \citenamefont {Wu}}]{Liu2015}%
  \BibitemOpen
  \bibfield  {author} {\bibinfo {author} {\bibfnamefont {J.}~\bibnamefont
  {Liu}}, \bibinfo {author} {\bibfnamefont {S.}~\bibnamefont {Liu}}, \bibinfo
  {author} {\bibfnamefont {N.}~\bibnamefont {Li}}, \bibinfo {author}
  {\bibfnamefont {B.}~\bibnamefont {Li}}, \ and\ \bibinfo {author}
  {\bibfnamefont {C.}~\bibnamefont {Wu}},\ }\href {\doibase
  10.1103/PhysRevE.91.042910} {\bibfield  {journal} {\bibinfo  {journal} {Phys.
  Rev. E}\ }\textbf {\bibinfo {volume} {91}},\ \bibinfo {pages} {042910}
  (\bibinfo {year} {2015})}\BibitemShut {NoStop}%
\bibitem [{\citenamefont {Dauxois}\ and\ \citenamefont
  {Peyrard}(2006)}]{Dauxois2006Physics}%
  \BibitemOpen
  \bibfield  {author} {\bibinfo {author} {\bibfnamefont {T.}~\bibnamefont
  {Dauxois}}\ and\ \bibinfo {author} {\bibfnamefont {M.}~\bibnamefont
  {Peyrard}},\ }\href@noop {} {\emph {\bibinfo {title} {Physics of solitons}}}\
  (\bibinfo  {publisher} {Cambridge University Press},\ \bibinfo {address}
  {Cambridge, UK},\ \bibinfo {year} {2006})\BibitemShut {NoStop}%
\bibitem [{\citenamefont {McMillan}(1976)}]{McMillan19761496}%
  \BibitemOpen
  \bibfield  {author} {\bibinfo {author} {\bibfnamefont {W.~L.}\ \bibnamefont
  {McMillan}},\ }\href {\doibase 10.1103/PhysRevB.14.1496} {\bibfield
  {journal} {\bibinfo  {journal} {Phys. Rev. B}\ }\textbf {\bibinfo {volume}
  {14}},\ \bibinfo {pages} {1496} (\bibinfo {year} {1976})}\BibitemShut
  {NoStop}%
\bibitem [{\citenamefont {Abe}\ and\ \citenamefont {Abe}(1979)}]{Abe1979}%
  \BibitemOpen
  \bibfield  {author} {\bibinfo {author} {\bibfnamefont {K.}~\bibnamefont
  {Abe}}\ and\ \bibinfo {author} {\bibfnamefont {T.}~\bibnamefont {Abe}},\
  }\href {\doibase 10.1063/1.862824} {\bibfield  {journal} {\bibinfo  {journal}
  {Phys. Fluids}\ }\textbf {\bibinfo {volume} {22}},\ \bibinfo {pages} {1644}
  (\bibinfo {year} {1979})}\BibitemShut {NoStop}%
\bibitem [{\citenamefont {Abe}\ and\ \citenamefont {Inoue}(1980)}]{ABE1980202}%
  \BibitemOpen
  \bibfield  {author} {\bibinfo {author} {\bibfnamefont {K.}~\bibnamefont
  {Abe}}\ and\ \bibinfo {author} {\bibfnamefont {O.}~\bibnamefont {Inoue}},\
  }\href {\doibase https://doi.org/10.1016/0021-9991(80)90105-9} {\bibfield
  {journal} {\bibinfo  {journal} {J. Comput. Phys.}\ }\textbf {\bibinfo
  {volume} {34}},\ \bibinfo {pages} {202 } (\bibinfo {year}
  {1980})}\BibitemShut {NoStop}%
\bibitem [{\citenamefont {Truskinovsky}\ and\ \citenamefont
  {Vainchtein}(2014)}]{Truskinovsky2014042903}%
  \BibitemOpen
  \bibfield  {author} {\bibinfo {author} {\bibfnamefont {L.}~\bibnamefont
  {Truskinovsky}}\ and\ \bibinfo {author} {\bibfnamefont {A.}~\bibnamefont
  {Vainchtein}},\ }\href {\doibase 10.1103/PhysRevE.90.042903} {\bibfield
  {journal} {\bibinfo  {journal} {Phys. Rev. E}\ }\textbf {\bibinfo {volume}
  {90}},\ \bibinfo {pages} {042903} (\bibinfo {year} {2014})}\BibitemShut
  {NoStop}%
\bibitem [{\citenamefont {Fermi}\ \emph {et~al.}()\citenamefont {Fermi},
  \citenamefont {Pasta},\ and\ \citenamefont {Ulam}}]{FermiStudies}%
  \BibitemOpen
  \bibfield  {author} {\bibinfo {author} {\bibfnamefont {E.}~\bibnamefont
  {Fermi}}, \bibinfo {author} {\bibfnamefont {J.}~\bibnamefont {Pasta}}, \ and\
  \bibinfo {author} {\bibfnamefont {S.}~\bibnamefont {Ulam}},\ }\href@noop {}
  {\enquote {\bibinfo {title} {{Studies of nonlinear problems: I}},}\ }\bibinfo
  {note} {{Los Alamos National Laboratory Report No. LA-1940, 1955; reprinted
  in Collected Papers of Enrico Fermi, edited by E. Segr{\'e} (University of
  Chicago Press, Chicago, 1965), Vol. 2, p 978}}\BibitemShut {NoStop}%
\bibitem [{\citenamefont {Zabusky}\ and\ \citenamefont
  {Kruskal}(1965)}]{Zabusky1965240}%
  \BibitemOpen
  \bibfield  {author} {\bibinfo {author} {\bibfnamefont {N.~J.}\ \bibnamefont
  {Zabusky}}\ and\ \bibinfo {author} {\bibfnamefont {M.~D.}\ \bibnamefont
  {Kruskal}},\ }\href {\doibase 10.1103/PhysRevLett.15.240} {\bibfield
  {journal} {\bibinfo  {journal} {Phys. Rev. Lett.}\ }\textbf {\bibinfo
  {volume} {15}},\ \bibinfo {pages} {240} (\bibinfo {year} {1965})}\BibitemShut
  {NoStop}%
\bibitem [{\citenamefont {Trillo}\ \emph {et~al.}(2016)\citenamefont {Trillo},
  \citenamefont {Deng}, \citenamefont {Biondini}, \citenamefont {Klein},
  \citenamefont {Clauss}, \citenamefont {Chabchoub},\ and\ \citenamefont
  {Onorato}}]{Trillo2016144102}%
  \BibitemOpen
  \bibfield  {author} {\bibinfo {author} {\bibfnamefont {S.}~\bibnamefont
  {Trillo}}, \bibinfo {author} {\bibfnamefont {G.}~\bibnamefont {Deng}},
  \bibinfo {author} {\bibfnamefont {G.}~\bibnamefont {Biondini}}, \bibinfo
  {author} {\bibfnamefont {M.}~\bibnamefont {Klein}}, \bibinfo {author}
  {\bibfnamefont {G.~F.}\ \bibnamefont {Clauss}}, \bibinfo {author}
  {\bibfnamefont {A.}~\bibnamefont {Chabchoub}}, \ and\ \bibinfo {author}
  {\bibfnamefont {M.}~\bibnamefont {Onorato}},\ }\href {\doibase
  10.1103/PhysRevLett.117.144102} {\bibfield  {journal} {\bibinfo  {journal}
  {Phys. Rev. Lett.}\ }\textbf {\bibinfo {volume} {117}},\ \bibinfo {pages}
  {144102} (\bibinfo {year} {2016})}\BibitemShut {NoStop}%
\bibitem [{\citenamefont {Costa}\ \emph {et~al.}(2014)\citenamefont {Costa},
  \citenamefont {Osborne}, \citenamefont {Resio}, \citenamefont {Alessio},
  \citenamefont {Chriv\`{\i}}, \citenamefont {Saggese}, \citenamefont
  {Bellomo},\ and\ \citenamefont {Long}}]{Costa2014108501}%
  \BibitemOpen
  \bibfield  {author} {\bibinfo {author} {\bibfnamefont {A.}~\bibnamefont
  {Costa}}, \bibinfo {author} {\bibfnamefont {A.~R.}\ \bibnamefont {Osborne}},
  \bibinfo {author} {\bibfnamefont {D.~T.}\ \bibnamefont {Resio}}, \bibinfo
  {author} {\bibfnamefont {S.}~\bibnamefont {Alessio}}, \bibinfo {author}
  {\bibfnamefont {E.}~\bibnamefont {Chriv\`{\i}}}, \bibinfo {author}
  {\bibfnamefont {E.}~\bibnamefont {Saggese}}, \bibinfo {author} {\bibfnamefont
  {K.}~\bibnamefont {Bellomo}}, \ and\ \bibinfo {author} {\bibfnamefont
  {C.~E.}\ \bibnamefont {Long}},\ }\href {\doibase
  10.1103/PhysRevLett.113.108501} {\bibfield  {journal} {\bibinfo  {journal}
  {Phys. Rev. Lett.}\ }\textbf {\bibinfo {volume} {113}},\ \bibinfo {pages}
  {108501} (\bibinfo {year} {2014})}\BibitemShut {NoStop}%
\bibitem [{\citenamefont {Friesecke}\ and\ \citenamefont
  {Wattis}(1994)}]{Friesecke1994}%
  \BibitemOpen
  \bibfield  {author} {\bibinfo {author} {\bibfnamefont {G.}~\bibnamefont
  {Friesecke}}\ and\ \bibinfo {author} {\bibfnamefont {J.~A.~D.}\ \bibnamefont
  {Wattis}},\ }\href {\doibase 10.1007/BF02099784} {\bibfield  {journal}
  {\bibinfo  {journal} {Commun. Math. Phys.}\ }\textbf {\bibinfo {volume}
  {161}},\ \bibinfo {pages} {391} (\bibinfo {year} {1994})}\BibitemShut
  {NoStop}%
\bibitem [{\citenamefont {Friesecke}\ and\ \citenamefont
  {Pego}(1999)}]{Friesecke1999}%
  \BibitemOpen
  \bibfield  {author} {\bibinfo {author} {\bibfnamefont {G.}~\bibnamefont
  {Friesecke}}\ and\ \bibinfo {author} {\bibfnamefont {R.~L.}\ \bibnamefont
  {Pego}},\ }\href {http://stacks.iop.org/0951-7715/12/i=6/a=311} {\bibfield
  {journal} {\bibinfo  {journal} {Nonlinearity}\ }\textbf {\bibinfo {volume}
  {12}},\ \bibinfo {pages} {1601} (\bibinfo {year} {1999})}\BibitemShut
  {NoStop}%
\bibitem [{\citenamefont {Friesecke}\ and\ \citenamefont
  {Matthies}(2002)}]{FRIESECKE2002211}%
  \BibitemOpen
  \bibfield  {author} {\bibinfo {author} {\bibfnamefont {G.}~\bibnamefont
  {Friesecke}}\ and\ \bibinfo {author} {\bibfnamefont {K.}~\bibnamefont
  {Matthies}},\ }\href {\doibase https://doi.org/10.1016/S0167-2789(02)00604-8}
  {\bibfield  {journal} {\bibinfo  {journal} {Physica D}\ }\textbf {\bibinfo
  {volume} {171}},\ \bibinfo {pages} {211 } (\bibinfo {year}
  {2002})}\BibitemShut {NoStop}%
\bibitem [{\citenamefont {Smets}\ and\ \citenamefont
  {Willem}(1997)}]{SMETS1997266}%
  \BibitemOpen
  \bibfield  {author} {\bibinfo {author} {\bibfnamefont {D.}~\bibnamefont
  {Smets}}\ and\ \bibinfo {author} {\bibfnamefont {M.}~\bibnamefont {Willem}},\
  }\href {\doibase https://doi.org/10.1006/jfan.1996.3121} {\bibfield
  {journal} {\bibinfo  {journal} {J. Funct. Anal.}\ }\textbf {\bibinfo {volume}
  {149}},\ \bibinfo {pages} {266 } (\bibinfo {year} {1997})}\BibitemShut
  {NoStop}%
\bibitem [{\citenamefont {Iooss}(2000)}]{Iooss2000}%
  \BibitemOpen
  \bibfield  {author} {\bibinfo {author} {\bibfnamefont {G.}~\bibnamefont
  {Iooss}},\ }\href {http://stacks.iop.org/0951-7715/13/i=3/a=319} {\bibfield
  {journal} {\bibinfo  {journal} {Nonlinearity}\ }\textbf {\bibinfo {volume}
  {13}},\ \bibinfo {pages} {849} (\bibinfo {year} {2000})}\BibitemShut
  {NoStop}%
\bibitem [{\citenamefont {Peyrard}\ \emph {et~al.}(1986)\citenamefont
  {Peyrard}, \citenamefont {Pnevmatikos},\ and\ \citenamefont
  {Flytzanis}}]{PEYRARD1986268}%
  \BibitemOpen
  \bibfield  {author} {\bibinfo {author} {\bibfnamefont {M.}~\bibnamefont
  {Peyrard}}, \bibinfo {author} {\bibfnamefont {S.}~\bibnamefont
  {Pnevmatikos}}, \ and\ \bibinfo {author} {\bibfnamefont {N.}~\bibnamefont
  {Flytzanis}},\ }\href {\doibase https://doi.org/10.1016/0167-2789(86)90023-0}
  {\bibfield  {journal} {\bibinfo  {journal} {Physica D}\ }\textbf {\bibinfo
  {volume} {19}},\ \bibinfo {pages} {268 } (\bibinfo {year}
  {1986})}\BibitemShut {NoStop}%
\bibitem [{\citenamefont {Flytzanis}\ \emph {et~al.}(1989)\citenamefont
  {Flytzanis}, \citenamefont {Pnevmatikos},\ and\ \citenamefont
  {Peyrard}}]{Flytzanis1989}%
  \BibitemOpen
  \bibfield  {author} {\bibinfo {author} {\bibfnamefont {N.}~\bibnamefont
  {Flytzanis}}, \bibinfo {author} {\bibfnamefont {S.}~\bibnamefont
  {Pnevmatikos}}, \ and\ \bibinfo {author} {\bibfnamefont {M.}~\bibnamefont
  {Peyrard}},\ }\href {http://stacks.iop.org/0305-4470/22/i=7/a=011} {\bibfield
   {journal} {\bibinfo  {journal} {J. Phys. A}\ }\textbf {\bibinfo {volume}
  {22}},\ \bibinfo {pages} {783} (\bibinfo {year} {1989})}\BibitemShut
  {NoStop}%
\bibitem [{\citenamefont {Kosevich}(1993)}]{Kosevich19932058}%
  \BibitemOpen
  \bibfield  {author} {\bibinfo {author} {\bibfnamefont {Y.~A.}\ \bibnamefont
  {Kosevich}},\ }\href {\doibase 10.1103/PhysRevLett.71.2058} {\bibfield
  {journal} {\bibinfo  {journal} {Phys. Rev. Lett.}\ }\textbf {\bibinfo
  {volume} {71}},\ \bibinfo {pages} {2058} (\bibinfo {year}
  {1993})}\BibitemShut {NoStop}%
\bibitem [{\citenamefont {Kosevich}\ \emph {et~al.}(2004)\citenamefont
  {Kosevich}, \citenamefont {Khomeriki},\ and\ \citenamefont
  {Ruffo}}]{Kosevich2004}%
  \BibitemOpen
  \bibfield  {author} {\bibinfo {author} {\bibfnamefont {Y.~A.}\ \bibnamefont
  {Kosevich}}, \bibinfo {author} {\bibfnamefont {R.}~\bibnamefont {Khomeriki}},
  \ and\ \bibinfo {author} {\bibfnamefont {S.}~\bibnamefont {Ruffo}},\ }\href
  {http://stacks.iop.org/0295-5075/66/i=1/a=021} {\bibfield  {journal}
  {\bibinfo  {journal} {Europhys. Lett.}\ }\textbf {\bibinfo {volume} {66}},\
  \bibinfo {pages} {21} (\bibinfo {year} {2004})}\BibitemShut {NoStop}%
\bibitem [{\citenamefont {Archilla}\ \emph {et~al.}(2015)\citenamefont
  {Archilla}, \citenamefont {Kosevich}, \citenamefont {Jim\'enez},
  \citenamefont {S\'anchez-Morcillo},\ and\ \citenamefont
  {Garc\'{\i}a-Raffi}}]{Archilla2015022912}%
  \BibitemOpen
  \bibfield  {author} {\bibinfo {author} {\bibfnamefont {J.~F.~R.}\
  \bibnamefont {Archilla}}, \bibinfo {author} {\bibfnamefont {Y.~A.}\
  \bibnamefont {Kosevich}}, \bibinfo {author} {\bibfnamefont {N.}~\bibnamefont
  {Jim\'enez}}, \bibinfo {author} {\bibfnamefont {V.~J.}\ \bibnamefont
  {S\'anchez-Morcillo}}, \ and\ \bibinfo {author} {\bibfnamefont {L.~M.}\
  \bibnamefont {Garc\'{\i}a-Raffi}},\ }\href {\doibase
  10.1103/PhysRevE.91.022912} {\bibfield  {journal} {\bibinfo  {journal} {Phys.
  Rev. E}\ }\textbf {\bibinfo {volume} {91}},\ \bibinfo {pages} {022912}
  (\bibinfo {year} {2015})}\BibitemShut {NoStop}%
\bibitem [{\citenamefont {Szeftel}\ \emph {et~al.}(1999)\citenamefont
  {Szeftel}, \citenamefont {Laurent-Gengoux},\ and\ \citenamefont
  {Ilisca}}]{Szeftel19993982}%
  \BibitemOpen
  \bibfield  {author} {\bibinfo {author} {\bibfnamefont {J.}~\bibnamefont
  {Szeftel}}, \bibinfo {author} {\bibfnamefont {P.}~\bibnamefont
  {Laurent-Gengoux}}, \ and\ \bibinfo {author} {\bibfnamefont {E.}~\bibnamefont
  {Ilisca}},\ }\href {\doibase 10.1103/PhysRevLett.83.3982} {\bibfield
  {journal} {\bibinfo  {journal} {Phys. Rev. Lett.}\ }\textbf {\bibinfo
  {volume} {83}},\ \bibinfo {pages} {3982} (\bibinfo {year}
  {1999})}\BibitemShut {NoStop}%
\bibitem [{\citenamefont {Szeftel}\ \emph {et~al.}(2000)\citenamefont
  {Szeftel}, \citenamefont {Laurent-Gengoux}, \citenamefont {Ilisca},\ and\
  \citenamefont {Hebbache}}]{SZEFTEL2000225}%
  \BibitemOpen
  \bibfield  {author} {\bibinfo {author} {\bibfnamefont {J.}~\bibnamefont
  {Szeftel}}, \bibinfo {author} {\bibfnamefont {P.}~\bibnamefont
  {Laurent-Gengoux}}, \bibinfo {author} {\bibfnamefont {E.}~\bibnamefont
  {Ilisca}}, \ and\ \bibinfo {author} {\bibfnamefont {M.}~\bibnamefont
  {Hebbache}},\ }\href {\doibase https://doi.org/10.1016/S0378-4371(00)00424-6}
  {\bibfield  {journal} {\bibinfo  {journal} {Physica A}\ }\textbf {\bibinfo
  {volume} {288}},\ \bibinfo {pages} {225 } (\bibinfo {year}
  {2000})}\BibitemShut {NoStop}%
\bibitem [{\citenamefont {Neogi}\ and\ \citenamefont
  {Mahan}(2008)}]{Neogi2008064306}%
  \BibitemOpen
  \bibfield  {author} {\bibinfo {author} {\bibfnamefont {S.}~\bibnamefont
  {Neogi}}\ and\ \bibinfo {author} {\bibfnamefont {G.~D.}\ \bibnamefont
  {Mahan}},\ }\href {\doibase 10.1103/PhysRevB.78.064306} {\bibfield  {journal}
  {\bibinfo  {journal} {Phys. Rev. B}\ }\textbf {\bibinfo {volume} {78}},\
  \bibinfo {pages} {064306} (\bibinfo {year} {2008})}\BibitemShut {NoStop}%
\bibitem [{\citenamefont {Jin}\ \emph {et~al.}(2010)\citenamefont {Jin},
  \citenamefont {Zhao},\ and\ \citenamefont {Hu}}]{Jin2010037601}%
  \BibitemOpen
  \bibfield  {author} {\bibinfo {author} {\bibfnamefont {T.}~\bibnamefont
  {Jin}}, \bibinfo {author} {\bibfnamefont {H.}~\bibnamefont {Zhao}}, \ and\
  \bibinfo {author} {\bibfnamefont {B.}~\bibnamefont {Hu}},\ }\href {\doibase
  10.1103/PhysRevE.81.037601} {\bibfield  {journal} {\bibinfo  {journal} {Phys.
  Rev. E}\ }\textbf {\bibinfo {volume} {81}},\ \bibinfo {pages} {037601}
  (\bibinfo {year} {2010})}\BibitemShut {NoStop}%
\bibitem [{\citenamefont {Chang}\ \emph {et~al.}(2006)\citenamefont {Chang},
  \citenamefont {Okawa}, \citenamefont {Majumdar},\ and\ \citenamefont
  {Zettl}}]{Chang06}%
  \BibitemOpen
  \bibfield  {author} {\bibinfo {author} {\bibfnamefont {C.~W.}\ \bibnamefont
  {Chang}}, \bibinfo {author} {\bibfnamefont {D.}~\bibnamefont {Okawa}},
  \bibinfo {author} {\bibfnamefont {A.}~\bibnamefont {Majumdar}}, \ and\
  \bibinfo {author} {\bibfnamefont {A.}~\bibnamefont {Zettl}},\ }\href@noop {}
  {\bibfield  {journal} {\bibinfo  {journal} {Science}\ }\textbf {\bibinfo
  {volume} {314}},\ \bibinfo {pages} {1121} (\bibinfo {year}
  {2006})}\BibitemShut {NoStop}%
\bibitem [{\citenamefont {Duncan}\ and\ \citenamefont
  {Wattis}(1992)}]{DUNCAN1992505}%
  \BibitemOpen
  \bibfield  {author} {\bibinfo {author} {\bibfnamefont {D.}~\bibnamefont
  {Duncan}}\ and\ \bibinfo {author} {\bibfnamefont {J.}~\bibnamefont
  {Wattis}},\ }\href {\doibase http://dx.doi.org/10.1016/0960-0779(92)90026-J}
  {\bibfield  {journal} {\bibinfo  {journal} {Chaos, Solitons and Fractals}\
  }\textbf {\bibinfo {volume} {2}},\ \bibinfo {pages} {505 } (\bibinfo {year}
  {1992})}\BibitemShut {NoStop}%
\bibitem [{\citenamefont {Ming}\ \emph {et~al.}(2016)\citenamefont {Ming},
  \citenamefont {Li},\ and\ \citenamefont {Ding}}]{Ming2016}%
  \BibitemOpen
  \bibfield  {author} {\bibinfo {author} {\bibfnamefont {Y.}~\bibnamefont
  {Ming}}, \bibinfo {author} {\bibfnamefont {H.-M.}\ \bibnamefont {Li}}, \ and\
  \bibinfo {author} {\bibfnamefont {Z.-J.}\ \bibnamefont {Ding}},\ }\href
  {\doibase 10.1103/PhysRevE.93.032127} {\bibfield  {journal} {\bibinfo
  {journal} {Phys. Rev. E}\ }\textbf {\bibinfo {volume} {93}},\ \bibinfo
  {pages} {032127} (\bibinfo {year} {2016})}\BibitemShut {NoStop}%
\bibitem [{\citenamefont {Zabusky}(1981)}]{ZABUSKY1981195}%
  \BibitemOpen
  \bibfield  {author} {\bibinfo {author} {\bibfnamefont {N.~J.}\ \bibnamefont
  {Zabusky}},\ }\href {\doibase https://doi.org/10.1016/0021-9991(81)90120-0}
  {\bibfield  {journal} {\bibinfo  {journal} {J. Comput. Phys.}\ }\textbf
  {\bibinfo {volume} {43}},\ \bibinfo {pages} {195 } (\bibinfo {year}
  {1981})}\BibitemShut {NoStop}%
\bibitem [{\citenamefont {Currie}\ \emph {et~al.}(1980)\citenamefont {Currie},
  \citenamefont {Krumhansl}, \citenamefont {Bishop},\ and\ \citenamefont
  {Trullinger}}]{Currie1980477}%
  \BibitemOpen
  \bibfield  {author} {\bibinfo {author} {\bibfnamefont {J.~F.}\ \bibnamefont
  {Currie}}, \bibinfo {author} {\bibfnamefont {J.~A.}\ \bibnamefont
  {Krumhansl}}, \bibinfo {author} {\bibfnamefont {A.~R.}\ \bibnamefont
  {Bishop}}, \ and\ \bibinfo {author} {\bibfnamefont {S.~E.}\ \bibnamefont
  {Trullinger}},\ }\href {\doibase 10.1103/PhysRevB.22.477} {\bibfield
  {journal} {\bibinfo  {journal} {Phys. Rev. B}\ }\textbf {\bibinfo {volume}
  {22}},\ \bibinfo {pages} {477} (\bibinfo {year} {1980})}\BibitemShut
  {NoStop}%
\bibitem [{\citenamefont {Theodorakopoulos}(1984)}]{Theodorakopoulos1984871}%
  \BibitemOpen
  \bibfield  {author} {\bibinfo {author} {\bibfnamefont {N.}~\bibnamefont
  {Theodorakopoulos}},\ }\href {\doibase 10.1103/PhysRevLett.53.871} {\bibfield
   {journal} {\bibinfo  {journal} {Phys. Rev. Lett.}\ }\textbf {\bibinfo
  {volume} {53}},\ \bibinfo {pages} {871} (\bibinfo {year} {1984})}\BibitemShut
  {NoStop}%
\bibitem [{\citenamefont {Fratalocchi}\ \emph {et~al.}(2008)\citenamefont
  {Fratalocchi}, \citenamefont {Conti}, \citenamefont {Ruocco},\ and\
  \citenamefont {Trillo}}]{Fratalocchi2008044101}%
  \BibitemOpen
  \bibfield  {author} {\bibinfo {author} {\bibfnamefont {A.}~\bibnamefont
  {Fratalocchi}}, \bibinfo {author} {\bibfnamefont {C.}~\bibnamefont {Conti}},
  \bibinfo {author} {\bibfnamefont {G.}~\bibnamefont {Ruocco}}, \ and\ \bibinfo
  {author} {\bibfnamefont {S.}~\bibnamefont {Trillo}},\ }\href {\doibase
  10.1103/PhysRevLett.101.044101} {\bibfield  {journal} {\bibinfo  {journal}
  {Phys. Rev. Lett.}\ }\textbf {\bibinfo {volume} {101}},\ \bibinfo {pages}
  {044101} (\bibinfo {year} {2008})}\BibitemShut {NoStop}%
\bibitem [{\citenamefont {Sievers}\ and\ \citenamefont
  {Takeno}(1988)}]{Sievers1988970}%
  \BibitemOpen
  \bibfield  {author} {\bibinfo {author} {\bibfnamefont {A.~J.}\ \bibnamefont
  {Sievers}}\ and\ \bibinfo {author} {\bibfnamefont {S.}~\bibnamefont
  {Takeno}},\ }\href {\doibase 10.1103/PhysRevLett.61.970} {\bibfield
  {journal} {\bibinfo  {journal} {Phys. Rev. Lett.}\ }\textbf {\bibinfo
  {volume} {61}},\ \bibinfo {pages} {970} (\bibinfo {year} {1988})}\BibitemShut
  {NoStop}%
\bibitem [{\citenamefont {Ming}\ \emph {et~al.}(2017)\citenamefont {Ming},
  \citenamefont {Ling}, \citenamefont {Li},\ and\ \citenamefont
  {Ding}}]{Ming2017}%
  \BibitemOpen
  \bibfield  {author} {\bibinfo {author} {\bibfnamefont {Y.}~\bibnamefont
  {Ming}}, \bibinfo {author} {\bibfnamefont {D.-B.}\ \bibnamefont {Ling}},
  \bibinfo {author} {\bibfnamefont {H.-M.}\ \bibnamefont {Li}}, \ and\ \bibinfo
  {author} {\bibfnamefont {Z.-J.}\ \bibnamefont {Ding}},\ }\href {\doibase
  10.1063/1.4985016} {\bibfield  {journal} {\bibinfo  {journal} {Chaos}\
  }\textbf {\bibinfo {volume} {27}},\ \bibinfo {pages} {063106} (\bibinfo
  {year} {2017})}\BibitemShut {NoStop}%
\bibitem [{\citenamefont {Lepri}(1998)}]{Lepri19987165}%
  \BibitemOpen
  \bibfield  {author} {\bibinfo {author} {\bibfnamefont {S.}~\bibnamefont
  {Lepri}},\ }\href {\doibase 10.1103/PhysRevE.58.7165} {\bibfield  {journal}
  {\bibinfo  {journal} {Phys. Rev. E}\ }\textbf {\bibinfo {volume} {58}},\
  \bibinfo {pages} {7165} (\bibinfo {year} {1998})}\BibitemShut {NoStop}%
\bibitem [{\citenamefont {Lee}\ \emph {et~al.}(2013)\citenamefont {Lee},
  \citenamefont {Kova{\v c}{\v c}},\ and\ \citenamefont {Cai}}]{Lee20133237}%
  \BibitemOpen
  \bibfield  {author} {\bibinfo {author} {\bibfnamefont {W.}~\bibnamefont
  {Lee}}, \bibinfo {author} {\bibfnamefont {G.}~\bibnamefont {Kova{\v c}{\v
  c}}}, \ and\ \bibinfo {author} {\bibfnamefont {D.}~\bibnamefont {Cai}},\
  }\href {\doibase 10.1073/pnas.1215325110} {\bibfield  {journal} {\bibinfo
  {journal} {Proc. Nat. Acad. Sci.}\ }\textbf {\bibinfo {volume} {110}},\
  \bibinfo {pages} {3237} (\bibinfo {year} {2013})}\BibitemShut {NoStop}%
\bibitem [{\citenamefont {Chen}\ and\ \citenamefont {Lee}(1979)}]{Chen1979264}%
  \BibitemOpen
  \bibfield  {author} {\bibinfo {author} {\bibfnamefont {H.~H.}\ \bibnamefont
  {Chen}}\ and\ \bibinfo {author} {\bibfnamefont {Y.~C.}\ \bibnamefont {Lee}},\
  }\href {\doibase 10.1103/PhysRevLett.43.264} {\bibfield  {journal} {\bibinfo
  {journal} {Phys. Rev. Lett.}\ }\textbf {\bibinfo {volume} {43}},\ \bibinfo
  {pages} {264} (\bibinfo {year} {1979})}\BibitemShut {NoStop}%
\bibitem [{\citenamefont {Rasinariu}\ \emph {et~al.}(1996)\citenamefont
  {Rasinariu}, \citenamefont {Sukhatme},\ and\ \citenamefont
  {Khare}}]{Rasinariu1996}%
  \BibitemOpen
  \bibfield  {author} {\bibinfo {author} {\bibfnamefont {C.}~\bibnamefont
  {Rasinariu}}, \bibinfo {author} {\bibfnamefont {U.}~\bibnamefont {Sukhatme}},
  \ and\ \bibinfo {author} {\bibfnamefont {A.}~\bibnamefont {Khare}},\ }\href
  {http://stacks.iop.org/0305-4470/29/i=8/a=027} {\bibfield  {journal}
  {\bibinfo  {journal} {J. Phys. A}\ }\textbf {\bibinfo {volume} {29}},\
  \bibinfo {pages} {1803} (\bibinfo {year} {1996})}\BibitemShut {NoStop}%
\bibitem [{\citenamefont {Matveev}(2002)}]{Matveev2002}%
  \BibitemOpen
  \bibfield  {author} {\bibinfo {author} {\bibfnamefont {V.~B.}\ \bibnamefont
  {Matveev}},\ }\href {\doibase 10.1023/A:1015149618529} {\bibfield  {journal}
  {\bibinfo  {journal} {Theor. Math. Phys.}\ }\textbf {\bibinfo {volume}
  {131}},\ \bibinfo {pages} {483} (\bibinfo {year} {2002})}\BibitemShut
  {NoStop}%
\end{thebibliography}
\end{document}